\documentclass[preprint,aps,pre,showpacs,epsfig,amsmath,amssymb,subfigure,eqsecnum]{revtex4}
\usepackage{epsfig,amsmath,amssymb,bm,epsf,graphics,graphicx,psfrag,verbatim,subfigure,framed}
\usepackage{bbm}
\usepackage{mathrsfs}
\usepackage{amsfonts}
\usepackage{color}

\newcommand{\rv}{{\vec r}}

\newcommand{\xv}{{\vec x}}

\newcommand{\yv}{{\vec y}}
\newcommand{\kv}{{\vec k}}
\newcommand{\Tr}{{\rm Tr}}

\newcommand{\be}{\begin{equation}}
\newcommand{\ee}{\end{equation}}
\newcommand{\ba}{\begin{eqnarray}}
\newcommand{\ea}{\end{eqnarray}}
\begin{document}
\title{Charge Renormalization and Charge Oscillation in Asymmetric Primitive Model}
\author{Mingnan Ding, Yihao Liang, Bing-Sui Lu, and Xiangjun Xing}
\affiliation{ Institute of Natural Sciences and Department of Physics and Astronomy,
Shanghai Jiao Tong University,
Shanghai, 200240 China
 \\
    \email{dmnphy@sjtu.edu.cn, xxing@sjtu.edu.cn}         
}

\date{\today} 


\begin{abstract} 
The Debye charging method is generalized to study the linear response properties of the asymmetric primitive model for electrolytes.  Analytic results are obtained for the effective charge distributions of constituent ions inside the electrolyte, from which all static linear response properties of system follow. It is found that, as the ion density increases, both the screening length and the dielectric constant receive substantial renormalization due to ionic correlations.  Furthermore, the valence of larger ion is substantially renormalized upwards by ionic correlations, whilst that of smaller ions remains approximately the same.  For sufficiently high density, the system exhibit charge oscillations. The threshold ion density for charge oscillation is much lower than the corresponding value for symmetric electrolytes.   Our results agree well with large scale Monte Carlo simulations. 

\end{abstract}

\maketitle

\section{Introduction}
It was pointed out by Kirkwood \cite{Kirkwood:1936ys} long ago that in sufficiently high densities, the ion-ion correlation functions of a symmetric electrolyte decays in an oscillatory fashion, a phenomenon called ``charge oscillation'' or ``charge ordering''.  Both analytic and numerical methods have been applied to study this phenomenon.  At the threshold of oscillation, the Debye length comparable with the ion diameter, and hence it was often argued that the mechanism of charge oscillation is the competition between hardcore repulsion and Coulombic attraction between opposite ions.  One-component plasma (OCP) also exhibits charge oscillation at high density \cite{Reviews:OCP}.  The underlying mechanism is the strong electrostatic repulsion between (likely-charged) ions, much like in dense neutral liquids.  Since OCP can be understood as the limit of extremely asymmetric electrolyte, where the valence of one component goes to zero, whilst symmetric electrolytes can be understood as the symmetric limit of asymmetric electrolytes, we would expect that charge oscillation also appears in asymmetric electrolytes, and the underlying mechanism is a combination of hard repulsion and the electrostatic repulsion between higher valence ions. 

Theoretically, the essence of charge oscillation can be captured by the {\em renormalized} electrostatic Green's function $G_{\! R}(\xv-\yv)$, which is defined as the mean potential at $\xv$, due to a  unit charge fixed at $\yv$ and all other screening ions.   The far field behaviors of all ion-ion pair correlations functions (which are more frequently used in liquid state physics) are identical to those of $G_{\! R}(\xv-\yv)$.  The Green's function however has the simplicity of satisfying a linear equation in the whole space:
\ba
\left( - \Delta + \alpha * \right) G_{\! R}(\xv - \yv)
 = \frac{1}{\epsilon} \delta(\xv - \yv).
 \label{LR-equation-1}
\ea
This equation was first derived by Kjellander and Mitchell \cite{Kjellander:1992nr,Kjellander:1994xy,Ennis:1995qf}, in the setting of ``dressed-ion theory''.   The kernel $\alpha(\xv - \yv)$ can be expressed in terms of charge-charge correlation functions \cite{DLX-ion-specific}. In this work, however, we shall not need this relation.   As shown by Kjellander and Mitchell, the Green's function decays in the form of a screened Coulomb potential in the far field:
\be
G_{\! R}(\xv-\yv) \sim \frac{e^{-\kappa_{\! R}|\xv - \yv|}}
{4 \pi \epsilon_{\! R}|\xv - \yv|}.  
\label{G-far-field}
\ee 
The parameters $\kappa_{\! R}$ and $\epsilon_{\! R}$ are determined by the pole structure of the Fourier transform $\hat{\alpha}(\kv)$, and are generically different from the dielectric constant of the pure solvent and the bare inverse Debye length as given in PB theory.  In the charge oscillation regime, $\kappa_{\! R}$ and $\epsilon_{\! R}$ become complex valued, and Eq.~(\ref{G-far-field})  should be understood as taking the real part. 

More remarkably,  Kjellander and Mitchell have also shown that the mean potential $ \phi_{\! \mu}$ due to a fixed constituent ion of specie $\mu$ satisfies an equation similar to Eq.~(\ref{LR-equation-1}), but with a renormalized charge distribution:
\ba
\left( - \Delta + \alpha * \right) \phi_{\! \mu}(\xv - \yv)
 = \frac{1}{\epsilon} K_{\! \mu}(\xv - \yv),
 \label{LR-equation-3}
\label{phi-K-G}
\ea
whose solution has the following far field asymptotics:
\be   
\phi_{\!\mu}(\xv-\yv) = 
\frac{q^{\rm R}_{ \mu} e^{- \kappa_{\! R}r}}
{4 \pi \epsilon_{\! R}}, 
\label{phi_mu-far}
\ee 
where $q^{\rm R}_{ \mu}$ plays the role of {\em renormalized charge} of the ion.   
The physical significance of  $K_{\!\mu}(\rv)$ is the {\em effective charge distribution} of ions of specie $\mu$.  There is an exact relation between the kernels $\alpha$ and all $K_{\!\mu}$'s: 
\begin{subequations}
\label{DIT-overall}
\be
\alpha = \frac{\beta}{\epsilon} \sum_{\mu} \bar{n}_{\mu}q_{\mu} K_{\!\mu},
\label{alpha-K-relation}
\ee  
where $q_{\mu}, \bar{n}_{\mu}$ are, respectively, the bare charge and bulk density of ions of specie $\mu$.  The Fourier transforms of $\alpha, K_{\!\mu}$ are related to $\kappa_{\! R}, \epsilon_{\! R} $ and $q^{\rm R}_{ \mu} $ via
\ba
\kappa_{\! R}^2 &=& \hat{\alpha}( \pm i \kappa_{\! R}),
\label{G_R-expansion-2-1}
 \\
 \frac{1}{2 i \varepsilon_{\! R} \kappa_{\! R}}
 &=& Res \! \left[ \hat{G}_{\!R}(k), i \kappa_{\! R} \right],
\label{G_R-expansion-2-2}
\\
q^{\rm R}_{ \mu}  &=& \hat{K}_{\! \mu}( i \kappa_{\! R} ). 
\label{G_R-expansion-2-3}
\ea
\label{G_R-expansion-2}
\end{subequations}
Setting $k = i \kappa_{\! R}$ in the Fourier transform of Eq.~(\ref{alpha-K-relation}), we find 
\be
\kappa_{\! R}^2 = 
\frac{\beta}{\epsilon} \sum_{\mu} 
\bar{n}_{\mu}q_{\mu} q^{\rm R}_{\! \mu}  . 
\label{alpha-kappa_R}
\ee
In the Poisson-Boltzmann theory, we have $K_{\!\mu} = q_{\mu} \delta(\rv)$, i.e., the effective charge distribution of an ion is just the bare charge distribution.  Hence $q^{\rm R}_{ \mu} = q_{\mu}, \varepsilon_{\! R} = \varepsilon$, and $\kappa_{\! R} = \kappa_0$, where
\be
  \kappa_0^2 = 
 \frac{\beta}{\epsilon} 
\sum_{\mu} \bar{n}_{\mu}q_{\mu}^2.
\label{alpha-kappa_0}
\ee  

Eqs.~(\ref{LR-equation-1}-\ref{alpha-kappa_0}) summarize the main results of ``dressed-ion theory'' due to Kjellander and Mitchell.  According to these relations, all linear response properties of the electrolyte, including renormalizations of charges, Debye length, as well as dielectric constant are encoded in the set of effective charge distributions $K_{\! \mu}$.  For reviews of the dressed ion theory, see \cite{DIT-review-Kjellander,DIT-review-2003}.  Note that our notations are different from those of Kjellander and Mitchell.  In reference \cite{DLX-ion-specific}, this theory is reformulated in a form that can describe ion-specific interactions.  

The main purpose of this work is to compute approximately the effective charge distributions $K_{\! \mu}$ in asymmetric primitive model.  We shall develop an analytic formalism for effective charge distribution of a generic charged hard sphere particle immersed in an electrolyte (Sec.~\ref{sec:formalism}).  By identifying this particle with a constituent ion, we obtain the effective charge distribution $K_{\!\mu}$ for each specie of ions, and further the linear response kernel $\alpha$, as well as various other parameters, e.g. $q^R_{\mu}, \kappa_{\! R}, \epsilon_{\! R}$.   As a special case, we shall apply the formalism to symmetric electrolytes, compute various renormalized parameters, and compare with previous theoretical results (Sec.~\ref{sec:sym-electrolyte}).  We shall also apply the formalism to asymmetric electrolyte (Sec.~\ref{sec:asym}).  In both cases, we compare our analytic results with large scale numerical simulations and find good agreements.  

\section{Analytic Formalism}
\label{sec:formalism}

\subsection{The Method of Debye Charging}

We shall study the {\em primitive model} of electrolytes, where the solvent is modeled implicitly as a homogeneous  media with a dielectric constant $\epsilon$, whilst ions are modeled as hard spheres with the {\em same} diameter $d$, and with a point charge at the center.   There is no {\em non-electrostatic} interactions other than the volume exclusion.  Furthermore, we assume that there is one specie of positive ion with charges $q_+ = m e$ and one specie of negative ion with charge $q_- = - n e$.  Hence $m,n$ are the valences of the positive and negative ions, whereas $e$ is the fundamental unit of electric charge.  Condition of charge neutrality then requires 
\be
m\, \rho_+ - n \, \rho_- = 0. \label{neutrality}
\ee
The Hamiltonian of a homogeneous unperturbed electrolyte is 
\ba
H_0  &=& \sum_{i < j} 
\Big[   q_i q_j \, G_0(\xv_{ij})
+ v_{\rm HC}\left( {|\xv_{ij}|}/{d}\right)
\Big],
\label{H-primitive}
\ea
where $q_i$ and $\xv_i$ are the charge and position of $i$-th constituent ion, $\xv_{ij} = \xv_i - \xv_j$ is the relative coordinate between $q_i, q_j$, and $G_0(\rv) = 1/4 \pi \epsilon \, r$ is the electrostatic Green's function in the bulk solvent, whilst the function $v_{\rm HC}(\xi)$ describes the hardcore interaction:
\be
v_{\rm HC}(\xi) = \left\{
\begin{array}{ll}
\infty,  \quad \quad
 & 0\leq \xi < 1; 
\vspace{2mm}\\
0, &\xi \geq 1.
\end{array}
\right.
\label{v_hs-def}
\ee
The canonical partition function and the Helmholtz free energy of the homogeneous electrolyte are given by
\begin{subequations}
\label{Z_N-F}
\ba
Z_0 &=& \Tr \,e^{ - \beta H_0 }
\equiv  \int \! \prod_{i=1}^N \! d^3 \xv_i
 \,e^{ - \beta H_0 }, \\
F_0 &=& - k_{\! B} T\, \log Z_0
=  - k_{\! B} T\, \log \Tr \,e^{ - \beta H_0 }. 
\label{Z_N-F-2}
\ea
\end{subequations}

\begin{figure}[t!]
	\centering
	\includegraphics[width=4.5 cm]{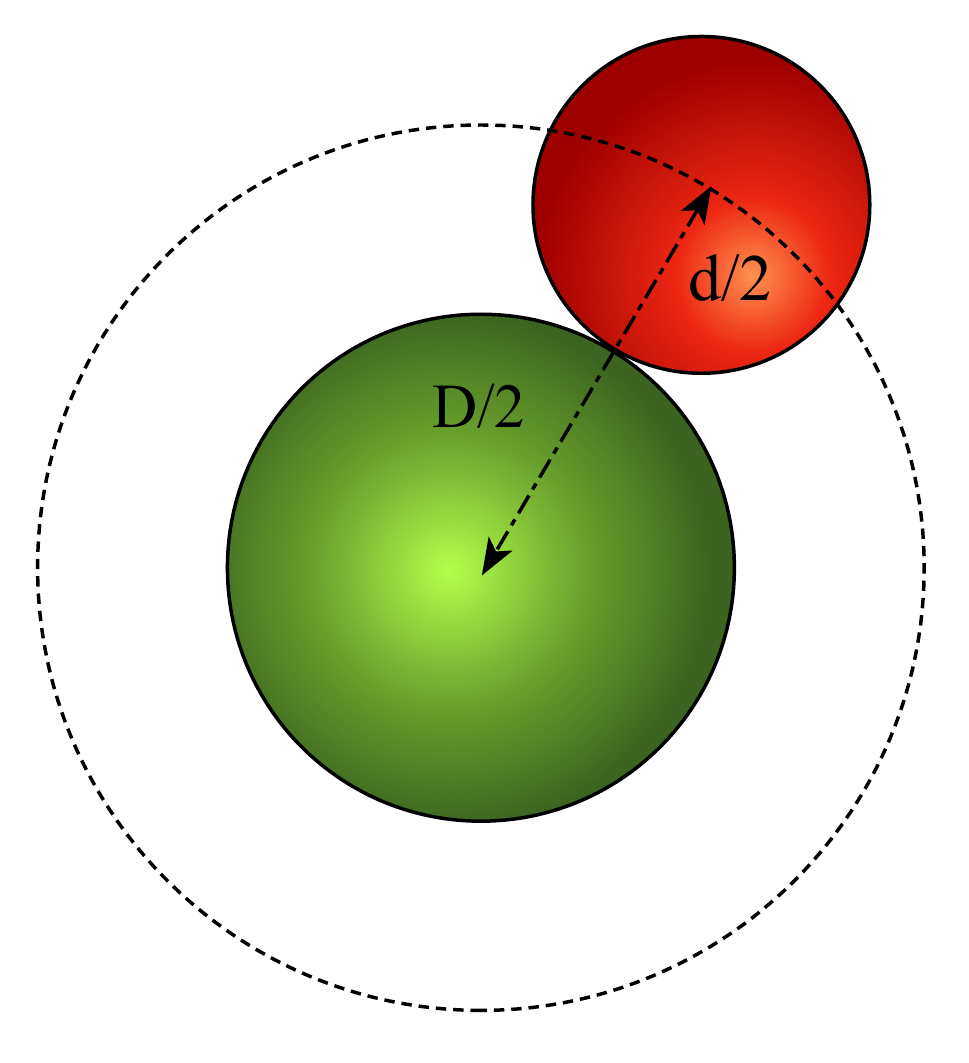}
	\caption{Insertion of a hard sphere (blue disk) with diameter $D$ creates a spherical excluded region for the centers-of-mass of all mobile ions, with a radius $R_c = (D + d)/2$.  The surface of the excluded region is called the {\em surface of contact}, schematically illustrated as the dashed circle. }
  \label{fig:hard-sphere-contact}
\vspace{-3mm}
\end{figure}


We shall perturb the homogeneous electrolyte by two means simultaneously: 1) Introducing an external potential $\phi^{\rm ex}$, which is sufficiently weak so that it can be treated using the linear response theory; 2) Inserting at $\rv$ a hard sphere particle with diameter $D$ and with a point charge $Q$ located at the center.  The center-of-mass coordinates of all mobile ions are consequently constrained outside a sphere with radius $R_c = (D+d)/2$.  This sphere shall be called {\em the contact surface}, and $R_c$ its radius. For an illustration, see Fig.~\ref{fig:hard-sphere-contact}. The total Hamiltonian of the perturbed system is then give by:
\ba
H &=& H_0 + \sum_{i} q_i \phi^{\rm ex}(\xv_i) 
+ Q \, \phi^{\rm ex}(\rv) + H_{\rm EP}. 
\label{H-inserted-ion}
\ea
The second and third terms in r.h.s. are, respectively, the interaction between the external potential $\phi^{\rm ex}$ and the electrolyte, and that between $\phi^{\rm ex}$ and the inserted particle.  The last term is the interaction between the inserted particle and all mobile ions: 
\be
H_{\rm EP} = \sum_i \Big[  Q  q_i G_0(\xv_i - \rv) + v_{\rm HC}(|\xv_i-\rv|/R_c). 
\Big] 
\ee
The change of free energy due to insertion of the particle is 
\ba
\Delta F(\rv,Q,R_c)
= & -& k_{\! B} T \, \log  \Tr \, 
e^{-\beta 
 \left( H_0+ \sum_{i} q_i \phi^{\rm ex}(\xv_i)
+ Q \phi^{\rm ex}(\rv)  + H_{\rm EP}
\right)}
\nonumber\\
&+&  k_{\! B} T \, \log  \Tr \, 
e^{-\beta 
 \left( H_0+ \sum_{i} q_i \phi^{\rm ex}(\xv_i)
\right)}
\nonumber\\
= & -& k_{\! B} T \log \frac{ \left\langle
e^{-\beta 
 \left(\sum_{i} q_i \phi^{\rm ex}(\xv_i)
+ Q \phi^{\rm ex}(\rv)  + H_{\rm EP}
\right)}
\right\rangle_0 }
{ \left\langle
e^{-\beta 
 \left(\sum_{i} q_i \phi^{\rm ex}(\xv_i)
\right)}
\right\rangle_0  },
\label{Delta-F-def}
\ea
where $\langle \, \cdot \, \rangle_0$ means average over the Gibbs distribution $e^{-\beta H_0}$ of the unperturbed electrolyte.  We shall call $\Delta F(\rv,Q,R_c)$  {\em the free energy of insertion}.   

In conjunction with the model system Eq.~(\ref{H-inserted-ion}), we shall also consider two other related systems: 
\begin{enumerate}
\item Electrolyte perturbed by $\phi^{\rm ex}$, but with no particle inserted:
\be
H = H_0 + \sum_{i} q_i \phi^{\rm ex}(\xv_i). 
\ee
All mobile ions in the electrolyte respond to the external potential, and create a total mean potential $\phi(\rv)$, which, according to the dressed-ion theory, satisfies a linear integro-differential equation [c.f. Eq.~(\ref{LR-equation-1})]:
\be
 - \Delta \phi(\rv) + \alpha * \phi (\rv)
   = \frac{1}{\epsilon} \rho^{\rm ex}_q(\rv),
   \label{phi-eqn}
\ee 
where $\rho^{\rm ex}(\rv) = - \epsilon \Delta \phi^{\rm ex}(\rv)$ is the external charge distribution that generates $\phi^{\rm ex}(\rv)$ in the first place.  

\item Electrolyte perturbed by the test particle, but with  $\phi^{\rm ex}(\rv)$ switched off:
\be
H =  H_0 + H_{\rm EP}. 
\ee
The corresponding free energy of insertion can be obtained from Eq.~(\ref{Delta-F-def}) by setting $\phi^{\rm ex}(\rv) = 0$:
\ba
 \lim_{\phi^{\rm ex} \rightarrow 0} \Delta F(\rv,Q,R_c)
  \rightarrow
-  k_{\! B} T \log \left\langle
e^{-\beta H_{\rm EP} }
\right\rangle_0. 
\label{Delta-F-def-0}
\ea
Note that it is independent of the location of insertion.  
\end{enumerate}

The difference between Eqs.~(\ref{Delta-F-def}) and (\ref{Delta-F-def-0}) is the {\em potential of mean force} (PMF) of the test particle, i.e., the effective interaction between the particle $Q$ and the external potential $\phi^{\rm ex}(\rv)$: 
\ba
U (\rv, Q, R_c) & = & \Delta F (\rv,Q,R_c) 
-  \lim_{\phi^{\rm ex} \rightarrow 0} \Delta F(\rv,Q,R_c)
 \label{U -Delta_F} \\
  & = & - k_{\! B} T \log \frac{ \left\langle
e^{-\beta 
 \left(\sum_{i} q_i \phi^{\rm ex}(\xv_i)
+ Q \phi^{\rm ex}(\rv)  + H_{\rm EP}
\right)}
\right\rangle_{\!\!0}  }
{ \left\langle
e^{-\beta 
 \sum_{i} q_i \phi^{\rm ex}(\xv_i)
}
\right\rangle_{\!0}  
\Big\langle
e^{-\beta H_{\rm EP} }
\Big\rangle_{\!\!0}} . 
\ea
If the potential $\phi^{\rm ex}(\rv)$ vanishes in the bulk (which is necessarily correct, if the external charge distribution $\rho^{\rm ex}(\rv)$ is localized), $U (\rv, Q; \phi)$ is also {\em the work needed to bring the particle from the bulk to the present position} $\xv$.   This is in fact the definition of PMF used in most literatures.

%

For a weak potential, $U(\xv, Q,R_c)$ can be expressed as a linear functional of $\phi^{\rm ex}$.  Actually, it is more useful to express $U(\xv, Q,R_c)$ in terms of the {\em total mean potential} $\phi$ prior to the insertion (which satisfies Eq.~(\ref{phi-eqn})):
\ba
U(\xv, Q,R_c) = \int_{\yv} K(\xv- \yv, Q) \phi(\yv). 
\label{U-K-def}
\ea
Note that $\phi$ and $\phi^{\rm ex}$ are proportional to each other in the linear regime.  The kernel $ K(\xv- \yv, Q)$ represents the {\em effective charge density} of the inserted particle.  
For the simple case of an infinitesimal point charge, the PMF is $U(\xv, dq,0) = dq\, \phi(\xv)$, which corresponds to an effective charge distribution $K(\xv - \yv) = dq\, \delta(\xv - \yv)$. 

To simplify the analysis, we shall choose the mean potential to be monochromatic $\phi(\rv) = \phi_0 \, e^{i \kv \cdot \rv}$.\footnote{which, of course, means that the external charge distribution $\rho^{\rm ex}(\rv)$ is monochromatic. }   Additionally, we shall also choose to insert the particle at the origin, so Eq.~(\ref{U-K-def}) becomes
\ba
U ({\bf 0}, Q,R_c) &=& \hat{K}(\kv, Q) \,
 \phi_0. 
\label{U_0-K_0}
\ea
Hence, all we have to do is to calculate the PMF $U ({\bf 0}, Q,R_c)$ of the test particle in a monochromatic background potential.  

Let us now take the derivative of Eq.~(\ref{Delta-F-def}) with respect to $Q$: 
\ba
\frac{\partial}{\partial Q} \Delta F(\rv,Q,R_c) 
&=& \frac{ \Tr
 \left[  \sum_i q_i G_0(\xv_i - \rv )
 + \phi^{\rm ex}(\rv) 
 \right] 
e^{- \beta H }
}
{
 \Tr \, 
e^{- \beta H }
}
\nonumber\\
&=&  \left\langle
 \sum_i q_i G_0(\xv_i - \rv)\right\rangle
 + \phi^{\rm ex}(\rv) 
\nonumber\\
&\equiv&  \psi(\rv, Q),
\label{DB-charging-1}
\ea
where the average $ \left\langle \, \cdot \, \right\rangle $ is defined with respect to the Gibbs measure $e^{-\beta H}$, with $H$ given by Eq.~(\ref{H-inserted-ion}).  Hence $\psi(\rv, Q)$ as defined is {\em the mean potential acting on $Q$ at the center of the test particle, due to all mobile charges $\{ q_i \}$ as well as the external charges $\rho^{\rm ex}(\rv)$.  }  Now if we switch off the external potential in Eq.~(\ref{DB-charging-1}) (and exchange the limit and derivative):
\ba 
\frac{\partial}{\partial Q}
\lim_{\phi^{\rm ex} \rightarrow 0} 
 \Delta F(\rv, Q, R_c)
= \lim_{\phi^{\rm ex} \rightarrow 0}  \psi(\rv, Q).
\label{dF-dq-2}
\ea
Subtracting this from Eq.~(\ref{DB-charging-1}) and using (\ref{U -Delta_F}),  we find:
\ba
\frac{\partial}{\partial Q} U  (\rv, Q, R_c)
&=&  \frac{\partial}{\partial Q}  \Delta F(\rv, Q) 
 - \frac{\partial}{\partial Q} 
 \lim_{\phi^{\rm ex} \rightarrow 0}  \Delta F(\rv, Q) 
 \nonumber\\
&=& \psi(\rv,Q) - \lim_{\phi^{\rm ex} \rightarrow 0} \psi(\rv,Q) 
\equiv \delta \psi(\rv,Q),
\label{delta-psi-def}
\ea
where $ \delta \psi(\rv,Q)$ is the difference between $\psi(\rv,Q)$ and its bulk value.  Following Debye \cite{Debye-charging}, we integrate Eq.~(\ref{delta-psi-def}) over $Q$, we find 
\begin{subequations}
\label{K-psi-integral}
\ba
U  (\rv, Q, R_c) =   U  (\rv, 0, R_c)
 +  \int_0^Q \!\! dq\, \delta \psi(\rv, q). 
\ea
\end{subequations}
Combining this result with Eq.~(\ref{U_0-K_0}), we see that $ \hat{K}(\kv,Q)$ can be calculated if we know $\psi({\bf 0},Q)$ (the mean potential acting on the test particle at the origin)  as well as $U  ({\bf 0}, 0, R_c)$, i.e., the PMF of a neutral particle.  We shall discuss $\psi(0,Q)$ in Sec.~\ref{sec:psi-0-q} and Sec.~\ref{sec:local}, and then discuss  $U  (\rv, 0, R_c)$ in Sec.~\ref{sec:charging-diameter}.


\subsection{Mean potential acting on the test particle}
\label{sec:psi-0-q}

To compute $\psi({\bf 0},Q)$, we shall first compute the total mean potential $\Phi(\rv, Q)$ at $\rv$, due to both the test ion fixed at the origin and all other mobile ions.  $\psi({\bf 0},Q)$ can then be obtained from $\Phi(\rv, Q)$ by subtracting off the Coulomb potential due to $Q$ itself and taking the local limit:
 \be
\psi({\bf 0},Q) 
  = \lim_{\rv \rightarrow 0} \left[ 
 \Phi(\rv,Q) - \frac{Q}{4 \pi \epsilon r} \right].
 \label{psi-phi}
\ee
Inside the contact surface $r < R_c$, no other ions can enter, and hence the potential $\Phi(\rv,Q)$ satisfies the Poisson equation:
\begin{subequations}
\be
-\epsilon \nabla^2 \Phi(\rv,Q)
 = Q \, \delta(\rv), \quad r< R_c.
\label{PB-nonlocal-2}
\ee
Outside the contact surface $r > R_c$, we shall use Eq.~(\ref{U-K-def})  to approximate the PMF of all constituent ions\footnote{This entails two assumptions: 1) that the linear approximation to the PMF is valid, and 2) that the effects of broken translational symmetry due to the boundary of the test particle can be ignored. }, so that the ion number density of specie $\mu$ is
\be
n_{\mu}(\rv) = \bar{n}_{\mu}\,e^{-\beta K_{\! \mu} * \Phi(\rv,Q)}.
\label{rho_mu-rv}
\ee
Consequently, $\Phi(\rv,Q)$ satisfies the following nonlinear (and nonlocal) partial integro-differential equation: 
\ba
-\epsilon \nabla^2 \Phi(\rv,Q)  &=& 
\sum_{\mu} q_{\mu}\, \bar{n}_{\mu}\, 
e^{ -\beta K_{\! \mu} * \Phi(\rv,Q)}
+ \rho_{\! q}^{\rm ex}(\rv), \quad  r >R_c.
\label{PB-nonlocal}
\ea 
\end{subequations}
Note that linearization of Eq.~(\ref{PB-nonlocal}) (together with Eq.~(\ref{alpha-K-relation})) leads to Eq.~(\ref{phi-eqn}). Eq.~(\ref{PB-nonlocal}) is an improvement over the nonlinear PBE, and reduces to the latter if one approximate $K_{\! \mu}$ by $q_{\mu} \delta(\rv)$.  
Additionally, $\Phi(\rv,Q)$ satisfies the standard electrostatic boundary conditions:  
\begin{subequations}
\label{BC-Phi}
\ba
&& \lim_{\rv \rightarrow \infty} \Phi(\rv,Q) = 0,  \\
&& \Phi(\rv,Q),  \frac{\partial }{\partial r} \Phi(\rv,Q) 
\,\, \mbox{continuous at} \,\, r = R_c. 
\ea
\end{subequations}
    
We shall calculate the PMF up to the second order in $Q$.  In view of Eq.~(\ref{K-psi-integral}), we only need to solve Eqs.~(\ref{PB-nonlocal}) to the first order in $Q$ and in $\phi_0$.  We decompose $\Phi(\rv, Q)$ into four parts:
\be
\Phi(\rv,Q) = \phi(\rv) + \phi^{\rm h.c.}(\rv)
+ Q \left[ G(\rv) +  \phi^{\rm c}(\rv) \right] 
 + O(Q^2) + O(\phi^2),
\label{Phi-decomp}
\ee
where $\phi(\rv)$ is the mean potential in the absence of the test ion, and satisfies the linear integro-differential equation (\ref{phi-eqn}).  $\phi^{\rm h.c.}(\rv)$ arises due to the insertion of a neutral hard sphere.  $G(\rv)$ is independent of $\phi$, whilst $\phi^{\rm c}$ is linear in $\phi$.  Both $G(\rv)$ and $\phi^{\rm c}$ are independent of $Q$.  All the ignored terms are at least quadratic either in $Q$ or in $\phi$. 

Let us set $Q = 0$ in Eq.~(\ref{Phi-decomp}), and obtain 
\be
\Phi(\rv,0) = \phi(\rv) + \phi^{\rm h.c.}(\rv) + O(\phi^2).
\label{Phi_0-decomp}
\ee
This satisfies Eqs.~(\ref{PB-nonlocal-2}) and (\ref{PB-nonlocal}) with $Q=0$, and corresponds to inserting a neutral hard sphere at the origin.  Expanding these equations to first order in $\phi$ and $\phi^{\rm h.c.}$, and subtracting off Eq.~(\ref{phi-eqn}), we find:
\begin{subequations}
\label{nonlocal-PDEs}
\ba
-  \Delta  \phi^{\rm h.c.}(\rv) = 
\left\{ \begin{array}{ll} 
- \alpha * \phi^{\rm h.c.}(\rv),
\quad
 & r >   R_c;
\vspace{2mm}\\
\,\,\,\, \alpha * \phi (\rv), & r < R_c . 
 \end{array} 
 \right. 
 \label{xi-eqn}
\ea

 The equation satisfied by $G(\rv)$ can be obtained by substituting Eq.~(\ref{Phi-decomp}) into Eqs.~(\ref{PB-nonlocal-2}) and (\ref{PB-nonlocal}) and extract the part that is linear in $Q$ and independent of $\phi$:
\ba
\left\{ \begin{array}{ll} 
-  \Delta  G (\rv) + \alpha * G(\rv)  = 0,
\quad
 & r >  R_c;
\vspace{2mm}\\
-  \Delta  G (\rv) =
\displaystyle{ \frac{1}{\epsilon} \, \delta(\rv) }
, & r < R_c . 
 \end{array} 
 \right. 
 \label{G-eqn}
\ea
One must be careful not identifying $G(\rv)$ with the renormalized Green's function $G_{\! R}(\rv)$ which satisfies  Eq.~(\ref{LR-equation-1}).   As one can see, $G(\rv)$ explicitly take into account the effects of hardcore repulsion, whilst $G_{\! R}(\rv)$ does not.  

Finally the equation satisfied by $\phi^{\rm c}$ can be obtained by extracting the bilinear term (proportional to $Q \, \phi$) of Eqs.~(\ref{PB-nonlocal-2}) and (\ref{PB-nonlocal}):
\ba
\left\{ \begin{array}{ll} 
-  \Delta  \phi^{\rm c} (\rv) + \alpha * \phi^{\rm c}(\rv)  
= {\epsilon}^{-1}  { \beta^2} 
 \sum_{\mu} n_{\mu} q_{\mu}
K_{\! \mu} \\
\quad\quad\quad\quad\quad\quad\quad\quad\quad
 * \left[ \phi(\rv) + \phi^{\rm h.c.}(\rv) \right]
K_{\! \mu}* G(\rv),
\quad
 & r >  R_c;
\vspace{2mm}\\
-  \Delta   \phi^{\rm c} (\rv) =0
, & r < R_c. 
 \end{array} 
 \right. 
\label{phi_c-eqn}
\ea
\end{subequations}
The boundary conditions for $ \phi^{\rm h.c.}(\rv)$, $G(\rv)$, and $\phi^{\rm c} (\rv)$ are identical to those for $\Phi(\rv, Q)$. 

\subsection{Local approximation}
\label{sec:local}

Eqs.~(\ref{nonlocal-PDEs}) are difficult to solve, because of the non-local nature of convolutions appearing in them.  To simplify the problem, we shall make the following {\em local} approximation for the kernel $\alpha$:
\begin{subequations}
\label{local-approx}
\be
\hat{\alpha}(\kv) = \kappa_{\! R}^2, \quad
{\alpha}(\xv) = \kappa_{\! R}^2\, \delta (\xv).
\label{local-approx-alpha}
\ee    
Correspondingly, the Green's function in Eqs.~(\ref{LR-equation-1}) are approximated by  
\ba
\hat{G}_{\! R}(\kv) \approx \frac{1}{k^2 + \kappa_{\! R}^2}, \quad
{G}_{\! R}(\rv) \approx \frac{1}{4 \pi \epsilon \, r} e^{- \kappa_{\! R} r}.
\label{local-approx-G}
\ea
In another word, this amounts to approximate the renormalized Green's function by a screened Coulomb potential with a renormalized Debye length.  Such an approximation is motivated by two considerations: 1) to preserve the large scale feature of renormalized theory outside the hard core, and 2) to make the analytical calculation feasible.  One can in principle make a more refined approximation, at a cost that the analyses can no longer be carried out explicitly.   To preserve the exact relation between $\alpha$ and $K_{\! \mu}$, Eq.~(\ref{alpha-K-relation}), we must make a similar approximation to 
the effective charge distributions $K_{\! \mu}$:
\be
\hat{K}_{\mu}(\kv) \approx q_{\! R}(q_{\mu}),
\quad
K_{\! \mu}(\rv) \approx q_{\! R}(q_{\mu}) \, \delta(\rv).
\label{local-approx-K}
\ee 
\end{subequations}
This amounts to simply replacing the bare charges by the renormalized charges, and ignoring the diffusive nature of the effective charge distributions.   Such an approximation turns out be rather successful, as we shall demonstrate below. 

With the local approximation, Eq.~(\ref{xi-eqn}) now reduces to 
\ba
-  \Delta  \phi^{\rm h.c.}(\rv) = 
\left\{ \begin{array}{ll} 
- \kappa_{\! R}^2 \,  \phi^{\rm h.c.}(\rv),
\quad
 & r >   R_c;
\vspace{2mm}\\
\,\,\,\, \kappa_{\! R}^2 \, \phi (\rv), & r <  R_c . 
 \end{array} 
 \right. 
 \label{xi-eqn-1}
\ea
Recall that $\phi(\rv) = \phi_0 \, e^{i \kv \cdot \rv}$  has the form of plane wave, and can be expanded in terms of spherical harmonics $Y_{lm}$ using the well-known formula:
\ba
 e^{i \kv \cdot \rv} &=& 4 \pi  \sum_{l=0}^{\infty} \sum_{m = -l}^l i^l  j_l(kr )
  \overline{Y_{lm}} (\hat{k}) Y_{lm} (\hat{r})   \nonumber\\
&=& j_0(k r) + {\rm anistropic},
\ea
where $j_l(k r)$ are the spherical Bessel functions of the first kind, and $\hat{k}, \hat{r}$ are the unit vectors parallel to $\kv, \rv$ respectively.  Likewise, $\phi^{\rm h.c.}(\rv)$ can also be expanded in terms of spherical harmonics.  To satisfy Eq.~(\ref{xi-eqn-1}) and be compatible with the boundary conditions at the origin and at the infinity, the expansion must have the following forms:
\be
\phi^{\rm h.c.}(\rv) = \left\{ 
\begin{array}{ll} 
\sum_{l,m} a_{l} \,
k_l(\kappa_{\! R} r) \,
 \overline{Y_{lm}} (\hat{k})
 Y_{lm}(\hat{r}),  \quad
& r > R_c; \vspace{3mm} \\
\sum_{l,m} \left[ 
c_{l}\, r^l + b_{l}(r) \right] 
 \overline{Y_{lm}} (\hat{k})
 Y_{lm}(\hat{r}), 
\quad & r <  R_c,
\end{array}
\right.
\label{phi_HC}
\ee
where $k_l(\kappa_{\! R} r)$ are the modified spherical Bessel functions of the second type, which vanish as $r \rightarrow \infty$, and $a_l,c_l, b_l(r)$ must be found.  Because $\phi^{\rm h.c.}(\rv) $ is continuous at the origin, all functions $b_{l}(r)$ with nonzero $l$ must vanish at $r = 0$.~\footnote{Alternatively, one may also show this by working out the solutions explicitly.}  We shall need two pieces of information about $ \phi^{\rm h.c.}(\rv) $ in this work: 1) $\phi^{\rm h.c.}(0) $ and 2) $\phi^{\rm h.c.}(\rv) $  averaged over the contact surface $r = R_c$.  For both quantities, the anisotropic channels with $l \neq 0$ make no contribution.  For the isotropic channel, $l=0$ and Eq.~(\ref{xi-eqn-1}) reduces to
\be
\left\{ 
\begin{array}{ll}
\begin{displaystyle}
- \left(  \frac{d^2}{dr^2} 
+ \frac{2}{r}\frac{d}{dr} \right)
\phi^{\rm h.c.}(r) + \kappa_{\! R}^2 \, 
\phi^{\rm h.c.}(r)  = 0,
\quad 
\end{displaystyle}
& r > R_c; \vspace{3mm}\\
\begin{displaystyle}
- \left(  \frac{d^2}{dr^2}
 + \frac{2}{r}\frac{d}{dr} \right)
\phi^{\rm h.c.}(r) 
=  \kappa_{\! R}^2 \,  \phi_0
j_0(kr),
\end{displaystyle}
& r< R_c. 
\end{array}
\right.
\ee 
The solution is (after imposing the boundary conditions Eq.~(\ref{BC-Phi}))
\be
\phi^{\rm h.c.}(r)  =
\left\{ 
\begin{array}{ll}
\begin{displaystyle}
 \phi_0\, f_1(kR_c, \kappa_{\! R}R_c) \, 
k_0(\kappa_{\! R} r),
\end{displaystyle}
\quad & r> R_c;
\vspace{3mm}\\
\begin{displaystyle}
\phi_0\left[  f_2 (kR_c, \kappa_{\! R}R_c)
+  \frac{\kappa_{\! R}^2}{k^2}
j_0(kr)
 \right], 
\end{displaystyle}
\quad & r <  R_c,
\end{array}
\right.
\label{phi_hc-sol}
\ee
where the functions $f_1 (x,y), f_2( x, y)$ are given by
\begin{subequations}
\label{psi_0-C-solution}
\ba
f_1 (x,y) &=& \frac{y^3 e^{y} ( \sin  x - x \cos x)}{x^3(1 +  y)}. 
\\
f_2( x, y) &=&  - \frac{y^2(y \sin  x+ x \cos x)}{x^3(1 + y)},
\ea
\end{subequations}

 Using the same approximation Eqs.~(\ref{local-approx}), Eq.~(\ref{G-eqn}) can be easily solved:
 \ba
G(r) = \left\{\begin{array}{ll}
\displaystyle{
\frac{e^{- \kappa_{\! R}(r- R_c)}}{4 \pi (1 + \kappa_{\! R}R_c) r},
}
& r > R_c; \vspace{3mm}\\
\displaystyle{
\frac{1}{4 \pi \epsilon r} 
+ \frac{\kappa_{\! R}}
{4 \pi \epsilon( 1+ \kappa_{\! R} R_c)}
}, \quad&  r < R_c. 
\end{array}\right.
 \label{G-eqn-1-sol}
\ea
Finally Eq.~(\ref{phi_c-eqn}) reduces to 
\ba
\left\{ \begin{array}{ll} 
\begin{displaystyle}
- \left(  \frac{d^2}{dr^2} 
+ \frac{2}{r}\frac{d}{dr} \right)
 \phi^{\rm c} (\rv) 
 + \kappa_{\! R}^2 \, \phi^{\rm c}(\rv)  
\end{displaystyle}  \\ \quad 
\begin{displaystyle}
=  {\epsilon}^{-1} {\beta^2} 
 \sum_{\alpha} \rho_{\alpha} q_{\alpha}
\left( q_{\! R}^{\alpha} \right)^2
 \left[ \phi(r) + \phi^{\rm h.c.}(r) \right] G(\rv)
, \end{displaystyle}
\quad
 & r >  R_c;
\vspace{2mm}\\
\begin{displaystyle}
- \left(  \frac{d^2}{dr^2} 
+ \frac{2}{r}\frac{d}{dr} \right) 
 \phi^{\rm c} (\rv) =0
, \end{displaystyle}
 & r < R_c. 
 \end{array} 
 \right. 
\label{phi_c-eqn-1}
\ea
Eq.~(\ref{phi_c-eqn-1}) will be solved in Sec.~\ref{sec:asym} for asymmetric electrolytes.  For symmetry electrolytes, $ \phi^{\rm c}(r) $ vanishes identically due to symmetry reason.  
    
\subsection{PMF of a neutral particle: ``contact value theorem''}
\label{sec:charging-diameter}
We shall now outline a method for  the PMF of a neutral hard sphere inside an electrolyte, Eq.~(\ref{U_0-K_0}).  As illustrated in Fig.~\ref{fig:hard-sphere-contact}, all ions are excluded from the region $r < R_c$.  Hence the partition function of the total system is
\ba
Z &=&  \int \!\! \prod_i  d^3 \rv_i \, \theta(r_i \! - \! R_c)
 \, e^{-\beta H }
=   \int_{R_c}^{\infty} \!\! \prod_i dr_i \!\! 
\int \!\! \prod_i d^2 \hat{\bf \rv}_i
 \, e^{-\beta H },
\ea
where $\hat{\bf \rv}_i$ is the unit vector parallel to ${\bf r}_i$.   The total free energy is:
\be
F = - k_{\! B} T \log {Z}
= F_0 + \Delta F({\bf 0} , 0, R_c). 
\label{contact-value-thm-0}
\ee
where $F_0$ is the free energy of the unperturbed electrolyte, whilst $\Delta F({\bf 0} , 0, R_c)$ is the free energy of insertion of the neutral hard sphere (c.f. Eq.~(\ref{Delta-F-def})).  Now let us take the differential of  $F$ with respect to $R_c$:
\ba
\frac{\partial \! F}{\partial \! R_c} d\! R_c 
&=& d\! R_c \frac{k_{\! B} T}{Z} \sum_i 
\int \!\!   d^2 \hat{r}_i \prod_{j, j\neq i} 
\int\! \!  d^3 \rv_j  \, \theta(r_j \! - \! R)
 \, \left( e^{-\beta H} \right)_{r_i = R}
\nonumber\\
&=& k_{\! B} T \, d\!  R_c \! \oint \! d^2 \hat{r} 
\Big\langle \sum_i\delta(\rv_i - \rv) \Big\rangle
\nonumber\\
&=& k_{\! B} T \, d\! R_c \,\oint \! d^2 \hat{r} 
\sum_{\mu} n_{\mu}(\rv) ,
\label{F-contact-derivative}
\label{contact-value-thm}
\ea
where $n_{\mu}(\rv)$ is the average ion number density of species $\mu$, and the integral $\oint d^2 \hat{r}$ is over the contact surface.  Eq.~(\ref{F-contact-derivative}) is a variation of the {\em contact value theorem} \cite{Henderson:1979kq}, which gives an exact relation between the particle number density for hard sphere systems and the pressure acting on a hard wall. It is important to note in Eq.~(\ref{F-contact-derivative}), we have used the fact that the Hamiltonian is independent of $R_c$.  This would not be correct if there is image charge effects.  Luckily enough, in the primitive model, the dielectric constants of the ions and of the solvent are identical, and hence image charge interactions do not appear.  

Now, according to Eq.~(\ref{rho_mu-rv}), the ion number density $n_{\mu}(\rv)$ (with a neutral particle fixed at the origin) is given by:
\be
n_{\mu}(\rv) = \bar{n}_{\mu}\,e^{-\beta K_{\! \mu} * \Phi(\rv,0)},
\label{rho_mu-rv-0}
\ee
where $\Phi(\rv,0)$ is given in Eq.~(\ref{Phi_0-decomp}).  Substituting this back into Eq.~(\ref{contact-value-thm}), linearizing in terms of $\Phi(\rv,0)$, using the local approximation Eq.~(\ref{local-approx-K}), and further integrating over the contact surface, we can express the r.h.s of Eq.~(\ref{contact-value-thm}) as a linear functional of $\Phi(\rv,0)$.  Finally keeping the part that is linear in $\Phi(\rv,0)$, and integrating over the radius of contact surface $R_c$, we obtain the PMF of a neutral hard sphere:
 \be
 U ({\bf 0}, 0, R_c) = - 4 \pi \sum_{\! \mu}
 \bar{n}_{\mu}\,q_{\mu}^R
 \int_0^{R_c} \!\!d R_c \, R_c^2
\langle \Phi(\rv,0) \rangle_{\rm cont},  
 \label{Delta_1-F}
 \ee
where $\langle \,\cdot\, \rangle_{\rm cont}$ means average over the contact surface.  We re-emphasize that this result is applicable only if the hard sphere has the same dielectric constant as the solvent. 

 

\section{Symmetric Electrolytes} 
\label{sec:sym-electrolyte}
Let us first apply the general method to the simple case of symmetric electrolytes, where positive/negatives ions have charges $\pm q$ and hard sphere diameter $d$.  The renormalized charges of positive and negative ions remain opposite to each other: $- q_{\! R}(- q)= q_{\! R}(q) \equiv q_{\! R}.$ Consequently the r.h.s. of (\ref{phi_c-eqn-1}) vanishes identically.  This means $\phi^{\rm c}$ vanishes identically, to the first order in $\phi$.  Likewise, the r.h.s. of Eq.~(\ref{Delta_1-F}) vanishes identically.  Hence $U ({\bf 0}, 0, R_c)  = 0$.

The total potential $\Phi(\rv, Q)$ can be obtained by substituting Eqs.~(\ref{phi_hc-sol}), (\ref{G-eqn-1-sol}) back into Eq.~(\ref{Phi-decomp}).  Using Eq.~(\ref{psi-phi}), we further calculate $\psi({\bf 0},Q)$, the mean potential acting on $Q$:
\ba
 \psi({\bf 0},Q) 
&=& \phi_0 \left[ 1+ \frac{\kappa_{\! R}^2}{k^2}   
+ f_2 (kR_c, \kappa_{\! R}R_c) \right]
+ \frac{Q \, \kappa_{\! R}}
{4 \pi \epsilon( 1 \! + \! \kappa_{\! R} R_c)}.
\ea
 We now use Eq.~(\ref{delta-psi-def}) to compute  $\delta  \psi({\bf 0},Q)$, and use Eq.~(\ref{K-psi-integral}) and the fact that $\hat{K}(\kv, 0)$ vanishes to obtain the effective charge distribution $\hat{K}(\kv, Q)$ for the test particle:
\be
\hat{K}(\kv, Q) = Q \left\{
1+ \frac{\kappa_{\! R}^2}{k^2}
-\frac{\kappa_{\! R}^2 \left[ \kappa_{\! R}  \sin (k R_c)+ k \cos
   (k R_c) \right]}{k^3 (1 + \kappa_{\! R} R_c )}
   \right\} + O(Q^3). 
\label{Khat-sym}
\ee
which is an entire function of $k$.  Recall that we have calculated $\hat{K}(\kv, Q) $ up to the order of $Q^2$, hence the ignored terms are at least of order $Q^3$. The renormalized charge can be obtained using Eq.~(\ref{G_R-expansion-2-3}):
\ba
{Q_{\! R}(Q)} = \hat{K}(i \kappa_{\! R}, Q) 
=  \frac{ {Q}  \,e^{\kappa_{\! R} R_c}}{1+ \kappa_{\! R} R_c}
+ O(Q^3).
\label{kappa_R-sym}
\ea
The fact that there is no term of order of $Q^2$ is actually enforced by the charge inversion symmetry: $Q_{\! R}(-Q) = - Q_{\! R}(Q)$. 

\subsection{Renormalized Debye length and renormalized dielectric constant}
We may apply Eq.~(\ref{Khat-sym}) to the constituent ions with $R_c = d, Q=q$, and further apply Eq.~(\ref{alpha-K-relation}) to calculate the linear response kernel $\alpha$:
\ba
\hat{\alpha}(\kv) 
&=& \kappa_0^2 \left[ 1+ \frac{\kappa_{\! R}^2}{k^2}
+ f_2 (kd, \kappa_{\! R}d) 
   \right], 
   \label{kappa_R-sym-1}
\ea
where $f_2(x,y)$ was already defined in Eqs.~(\ref{psi_0-C-solution}). 

The renormalized Debye length can be obtained via Eqs.~(\ref{DIT-overall}) and (\ref{kappa_R-sym}):
\begin{subequations}
\label{results-sym}
\be
\left( \! \frac{\kappa_{\! R}}{\kappa_0} \!\right)^2 
=  \frac{e^{\kappa_{\! R} d}}{1+ \kappa_{\! R} d} 
=\frac{q_{\! R}}{q}, 
\label{kappa_R-sym-1}
\ee
where $q_{\! R}$ is the renormalized charge of the positive constituent ion. This is a self-consistent equation for the renormalized inverse Debye length $\kappa_{\! R}$.   We can also calculate the renormalize dielectric constant:
\ba
\frac{\epsilon_{\!R}}{\epsilon}  =   
 2 &-& \frac{1}{2}   \kappa_{\! R} d
 - e^{- \kappa_{\! R} d} \left[ (1+  \kappa_{\! R} d) 
 - \frac{ 1}{2}  \sinh \kappa_{\! R} d  \right].
 \quad 
\label{ep_R-sym} 
\ea
\end{subequations}

\subsection{Charge Oscillation}
  Careful analysis of Eq.~(\ref{kappa_R-sym-1}) indicates a critical value $ \kappa_0^*$ defined by
\be
 \kappa_0^* d = \sqrt{2} \left(2+\sqrt{3}\right)
   e^{-\frac{1}{2}-\frac{\sqrt{3}}{2}}
   \approx 1.3465,
\ee
such that for $ \kappa_0 > \kappa_0^*$, there is no real root for Eq.~(\ref{kappa_R-sym-1}).  What happens is that a pair of real roots collide with each other at $ \kappa_0^*$ and bifurcate into the complex plane.  Consequently the renormalized charge $q_{R}$ and renormalized dielectric constant $\epsilon_{\!R}$ also become complex valued.  Whilst our local approximation is not really applicable if $\kappa_0$ is sufficiently close to $\kappa_0^*$,  it seems very natural to argue that one should take the real parts of Eq..~(\ref{G-far-field}) and (\ref{phi_mu-far}) if the relevant parameters become complex.  Therefore in the regime $ \kappa_0 > \kappa_0^*$, mean potential decays in an oscillatory fashion and the system exhibits {\em charge oscillation}.  The corresponding $\kappa_{\! R}^*$ at the threshold is
\begin{subequations}
\ba
\kappa_{\! R}^* d &=& 1 + \sqrt{3} \approx 2.732.
\ea  
\end{subequations}

\subsection{Comparison with simulations}

\begin{figure*}[t]
	\centering
\subfigure[]{	
	\includegraphics[width=5.5cm]{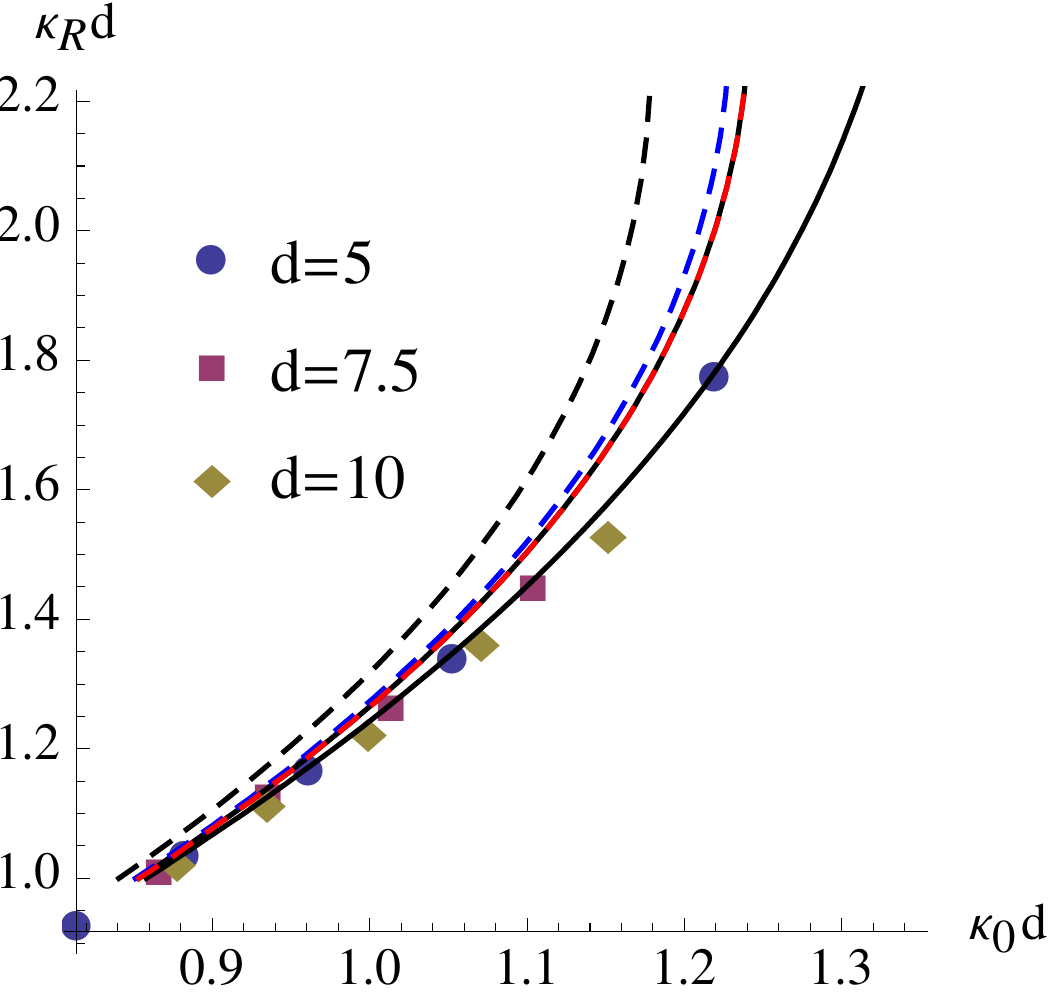}
}
\subfigure[]{	
	\includegraphics[width=6.8cm]{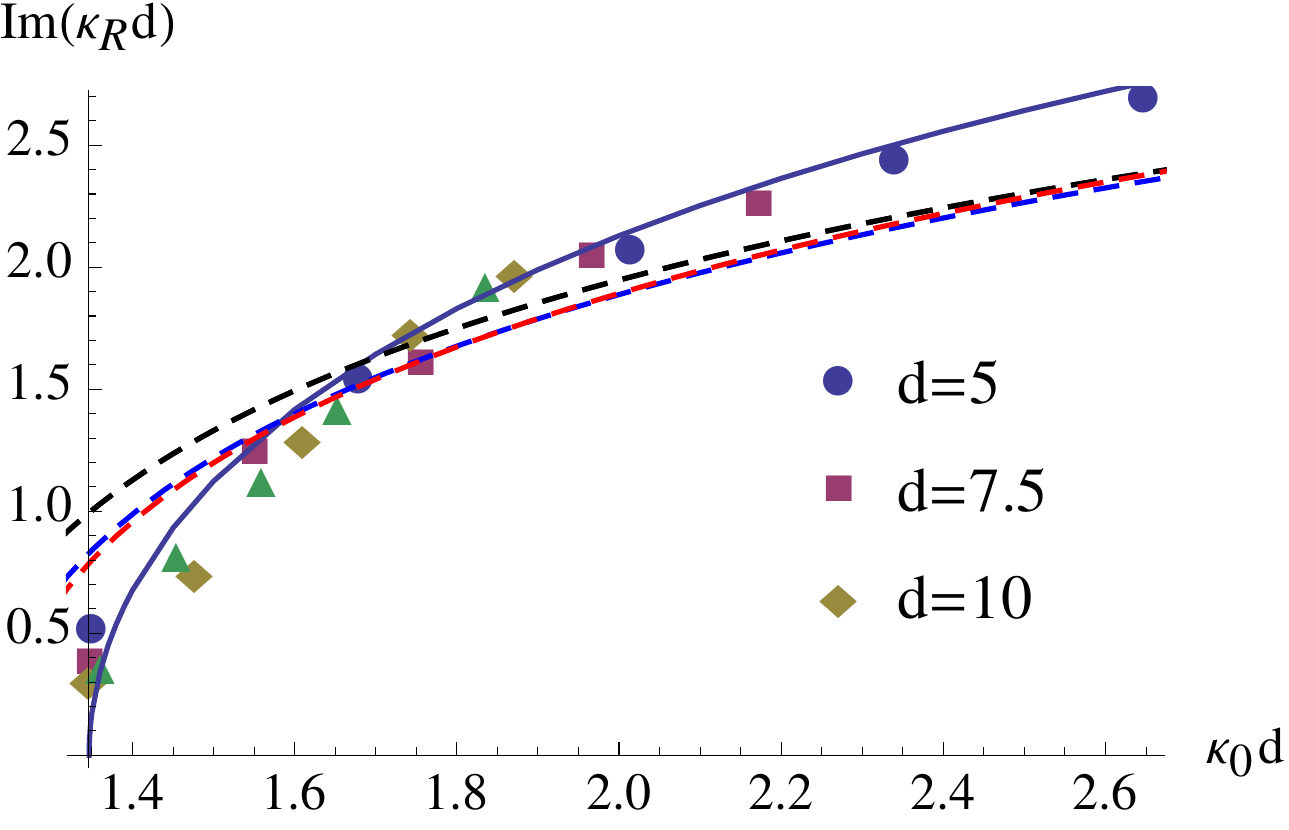}
}
\subfigure[]{	
	\includegraphics[width=6cm]{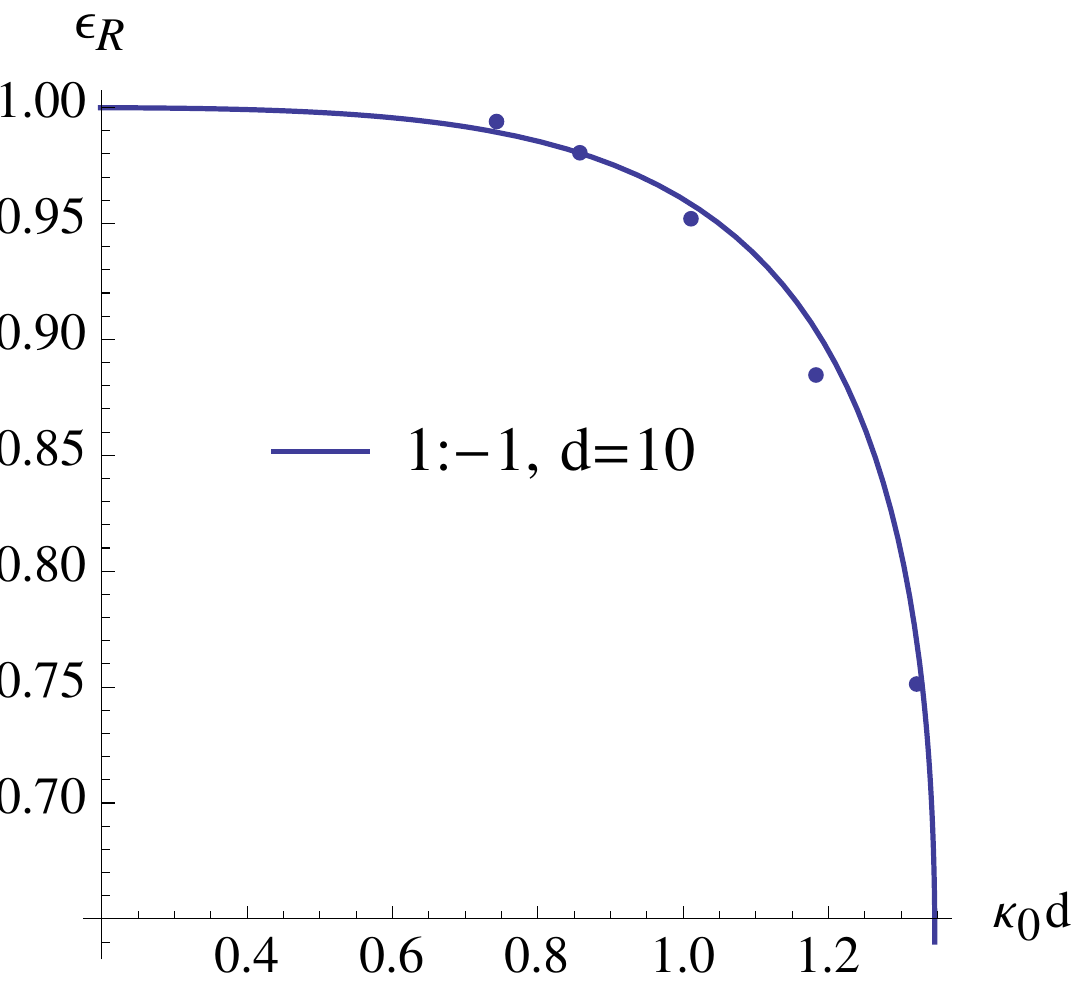}
}
\vspace{-3mm}
	\caption{{\bf Renormalized parameters for dense symmetric electrolyte.} Left: Renormalized v.s. bare inverse Debye length. Middle: the imaginary part of $\kappa_{\! R} d$ in the charge oscillation regime.   Symbols: MC simulation results.  Solid curves: our analytic results Eq.~(\ref{kappa_R-sym}).    The dashed curves are, respectively: black, generalized Debye-Huckel by Lee and Fisher \cite{lee1997charge}; red and blue: results using other approaches, LMPB and MSA, both from reference \cite{Ennis:1995qf}.  The other popular theory HNC does not yield a close form result. Right: Renormalized dielectric constant, for which we have not found any previous analytic result. Simulations details are discussed in a separate publication \cite{DXL-GPU}.  
	\vspace{-3mm}
 }
  \label{fig:R-debye-density-1-1}
\end{figure*}

We simulated $1:-1$ electrolytes with three different ion sizes: $d = 5\AA, 7.5 \AA, 10 \AA$ respectively, and determine all renormalized parameters. The simulation method is described elsewhere. \cite{DXL-GPU}  As shown in Fig.~\ref{fig:R-debye-density-1-1}, our MC results seem to agree with Eqs.~(\ref{results-sym}) remarkably well, both below and above threshold of charge oscillation.  A general argument due to Kjellander and Mitchell shows that ${\epsilon_{\! R}^*}$ should vanish at the threshold.  This contradicts our result Eq.~(\ref{ep_R-sym}), which gives $\epsilon_{\! R}^* \approx 0.64$.   We note that near the threshold, the local approximation Eq.~(\ref{local-approx-G}) breaks down, and therefore  Eq.~(\ref{ep_R-sym}) can not be trusted.  In any case, simulations near the threshold are very difficult and we have no reliable results to report so far.  



\subsection{Effective charge distribution and mean potential}

The real space version of the effective charge distribution can also be calculated, by Fourier transforming Eq.~(\ref{Khat-sym}):
\ba
K(\rv, Q) =Q \, \delta(\rv) +
Q \frac{\kappa_{\! R}^2 (\kappa_{\! R} R_c
  - \kappa_{\! R}  r+1)}{4 \pi  
 (1 + \kappa_{\! R} R_c ) r} \theta(R_c - r),
 \label{K-sym}
\ea
where $\theta(R_c - r)$ is the Heaviside step function.  The first term (Dirac delta function) is clearly due to the bare charge of the test particle.  The second term is positive and monotonically decreasing, and vanishes identically outside the contact surface ($r >R_c$). It is the diffusive part due to charge correlations.  Note that even though the second term is singular (diverges as $r^{-1}$) at the origin, it does not generates any singularity in the mean potential.  The mean potential, which is related to $\hat{K}(\kv, Q)$ via Eq.~(\ref{phi-K-G}), can also be explicitly calculated:
\ba
\phi(\rv, Q) = \left\{ \begin{array}{ll}
\begin{displaystyle}
\frac{Q_{\! R}\, e^{-\kappa_{\! R} r}}{4 \pi \epsilon r}, 
\end{displaystyle}
& r> R_c; 
\vspace{3mm}\\
\begin{displaystyle}
\frac{Q}{4 \pi \epsilon r} 
- \frac{Q \kappa_{\! R}  }
{4 \pi \epsilon (1+ \kappa_{\! R} R_c)},
\end{displaystyle} \quad\quad
& 0< r< R_c,  
\end{array}
\right.
\ea
which has the exact form of screened Coulomb potential outside the contact surface.   Since $Q_{\! R}$ is given by Eq.~(\ref{kappa_R-sym}) and always has the same sign as the bare charge, we see that there is {\em no} charge inversion in symmetry electrolytes at the level of our approximation.  

The results shown in this section were also obtained by Kjellander \cite{Kjellander:1995fk} some time ago.  We rederive these results to illustrate the method of Debye charging.  

\section{Asymmetric Electrolytes}
\label{sec:asym}
Let us now consider asymmetric electrolytes.  Let the renormalized charges of constituent ions be, respectively, $q_+^R = m_{\! R} e$, and $q_-^R = n_{\! R}$, where $m_{\! R} = q_+^R/e$, $n_{\! R} = q_-^R/e$ are the renormalized valences.  Using the neutrality condition (\ref{neutrality}) in Eqs.~(\ref{alpha-kappa_0}) and (\ref{alpha-kappa_R}) we can obtain 
\begin{subequations}
\ba
\kappa_0^2 &=& \epsilon^{-1} \beta  e^2 \rho_+ m (m+n)
, \\
\kappa_{\! R}^2 &=& \epsilon^{-1} \beta  e^2 
\rho_+ m (m_{\! R} + n_{\! R}). 
\ea
\label{kappaR-kappa0}
\end{subequations}
Dividing, we find a useful relation between  $\kappa_{\! R}$ and the renormalized valences: 
\be
\left( \! \frac{\kappa_{\! R}}{\kappa_0} \!\right)^2 
= \frac{m_{\! R} + n_{\! R}}{m+n}. 
\label{kappaR-kappa}
\ee

\subsection{PMF of neutral hard sphere continued}
Let us calculate the PMF of a neutral hard sphere $U(0,0,R_c)$ using Eq.~(\ref{Delta_1-F}).  For this purpose, we need the mean potential $\Phi(\rv, 0)$ in the presence of a neutral hard sphere, Eq.~(\ref{Phi_0-decomp}), with $\phi^{\rm h.c.}(\rv)$ given by Eq.~(\ref{phi_HC}).  
Averaging over the contact surface is trivial, because we have already thrown out the anisotropic parts.  The result is
\be
 \left \langle \Phi(\rv,0) \right \rangle_{\rm cont}
= \phi_0 \big[ 
 j_0(k R_c)
+ f_1(k R_c, \kappa_{\! R} R_c) 
k_0(\kappa_{\! R} R_c)
\big]. 
\ee
Substituting this into Eq.~(\ref{Delta_1-F}) and further back into Eq.~(\ref{U_0-K_0}), we find the effective charge distribution $\hat{K}(\kv, 0)$ for a neutral hard sphere: 
\ba
\hat{K}(\kv, 0) &=& 
-  4 \pi  R_c^3 \left( \sum_{\alpha} \rho_{\alpha} q_{\alpha}^{R}\right)
\Psi_0(k R_c, \kappa_{\! R} R_c). 
\label{K-hat-0}
\ea
where the function $\Psi_0(x, y)$ is defined as
\ba
\Psi_0(x, y)
&=& \int_0^1  \big[ \, j_0(x t)
+ f_1(x t, y t) \, k_0(yt) \, \big]  t^2\, dt
\nonumber\\
&=&
\frac{1}{x^4 y} 
\Big[ -x \, \text{Ci} \left(x +{x}{y}^{-1}\right) 
\left(x \cos (x y^{-1}) - y \sin \left( x y^{-1}\right)\right)
\nonumber\\
&+& x \, \text{Ci} \left( x y^{-1} \right) 
\left( x \cos \left( x y^{-1}\right)
-y \sin \left( x y^{-1}\right)\right)
\nonumber\\
&-& x\,  \text{Si}\left(x + x y^{-1} \right) 
\left(x \sin \left(x y^{-1}\right)
+y \cos \left(x y^{-1}\right)\right)
\nonumber\\
&+& x\,  \text{Si}\left(x y^{-1}\right) 
\left(x \sin \left(x y^{-1}\right)
+y \cos \left(x y^{-1}\right)\right)
\nonumber\\
&-& y \left(\left(x^2+2 y\right) \cos (x)+x (y-2)
   \sin (x)\right)+2 y^2
   \Big], 
   \label{Psi_0-def}
\ea
where $\text{Ci}(x)$ and $\text{Si}(x)$ are the cosine integral and sine integral functions: 
\ba
\text{Ci}(z) &=& - \int_z^{\infty} \frac{\cos t}{t} {dt}, 
\quad
\text{Si}(z) =  \int_0^z \frac{\sin t}{t} {dt}. 
\ea
Both $\text{Ci}(z)$ and $\text{Si}(z)$, and hence  $\Psi_0(k R_c, \kappa_{\! R} R_c)$ as well, are entire functions.  

Using Eqs.~(\ref{neutrality}) and (\ref{kappaR-kappa0}), we can rewrite Eq.~(\ref{K-hat-0}) into the following dimensionless form:
\ba
\frac{1}{e} \hat{K}(\kv, 0) &=& - \frac{\kappa_0^2 R_c^3}{b}
\frac{\left(m_Rn - n_R m \right)}{mn(m+n)}
\Psi_0(kR_c, \kappa_{\! R}R_c),
 \label{khat_0-result}
\ea
where $b = e^2/4 \pi \epsilon T$ is the Bjerrum length.  If both the inserted particle and the constituent ions are point-like, $R_c \rightarrow 0$, and $\Psi_0(0,0) = 1/3$, and hence 
\be
\hat{K}(\kv, 0 )   \rightarrow
- \frac{4 \pi}{3} R_c^3   
\sum_{\mu} \bar{n}_{\mu}q_{\mu}^{R}. 
\ee

\subsection{The correlation potential $\phi^{\rm c}(\rv)$}

\label{sec:asymmetric} 
To calculate $\phi^{\rm c}(0)$, we only need to solve the isotopic channel of Eq.~(\ref{phi_c-eqn-1}).   The isotropic component of r.h.s. of Eq.~(\ref{phi_c-eqn-1}) can be easily shown as 
\ba
&& \frac{ \phi_0 Q}{e } \theta(r - R_c) (m_{\! R} - n_{\! R})
 \kappa_{\! R}^2 (\kappa_{\! R} b) 
 \frac{e^{ \kappa_{\! R}R_c}}{1+  \kappa_{\! R}R_c}
\nonumber\\
&\times& 
  k_0(\kappa_{\! R} r)
\big[  j_0(k r)
+ f_1(kR_c, \kappa_{\! R}R_c) k_0(\kappa_{\! R} r)
\big].  
\ea  
where $f_1(x,y)$ is defined in Eqs.~(\ref{psi_0-C-solution}), and  $ j_0(u) = {\sin u}/{u}, \quad k_0 (u) = {e^{-u}}/{u}$.    $\phi^{\rm c}(\rv)$ can now be found using standard Liouville method \cite{Stone-Math}:
\ba
\phi^{\rm c}(0)
&=& \frac{2\, \phi_0 Q}{e}\,  
(m_{\! R} - n_{\! R}) 
\,(\kappa_{\! R} b)\, \Psi_2 (kR_c, \kappa_{\! R} R_c),   
    \label{Psi_2-def}\\
\Psi_2 (x, y) &\equiv&  \frac{e^{2 y} } {4(1+y)^2 } 
\Big[ 2 f_1(x, y) \, {\rm E}_1 \! \left( 3 y \right)
\nonumber\\  
&+& i \left( \frac{y}{x} \right)
 \big({\rm E}_1 \! \left( 2 y + i x\right)
   -{\rm E}_1 \! \left( 2 y-i x \right)\big)\Big],
    \label{Psi_2-def-2}
\ea   
where is $ {\rm E}_1(z)$ the {\em exponential integral function} \cite{HMF-Milton-Stegun}, defined as 
\ba
 {\rm E}_1(z) = \int_{z}^{\infty}  t^{-1} e^{-t}  dt. 
\ea 
$ {\rm E}_1(z)$ has a logarithmic singularity at the origin $z = 0$, and a branch cut on the negative real axis.  Consequently, the function $\Psi_2(x, y)$ as a function of complex variable $x$ has two branch cuts on the imaginary axis: one from $2 i y$ to $i \infty$, and the other from $-2 i y$ to $-i \infty$.  These singularities have no influence on the leading order asymptotics of mean potential by a charged hard sphere, or on the interaction between two charged spheres, as the leading order asymptotics of the latter quantities are controlled by the pole $k = i \kappa_{\! R}$, where the function $\Psi_2(k R_c, \kappa_{\! R}R_c)$ is analytic.  

Substituting this and Eqs.~(\ref{phi_hc-sol}), (\ref{G-eqn-1-sol}) back into Eqs.~(\ref{Phi-decomp}) and (\ref{psi-phi}), we find the potential acting on the the charge $Q$ (subtracting its bulk value):
\ba
\delta \psi({\bf 0},Q) &=& \phi_0 \bigg[
 1+ \frac{\kappa_{\! R}^2}{k^2}   
+ f_2 (kR_c, \kappa_{\! R}R_c)
\nonumber\\  
&+& \frac{2 Q}{e}\,  (m_{\! R} - n_{\! R})\,
(\kappa_{\! R} b)\,\Psi_2 (kR_c, \kappa_{\! R} R_c) 
\bigg].  
\ea
Now using Eqs.~(\ref{K-psi-integral}) and (\ref{khat_0-result}) to carry out the Debye charging process, we finally obtain the effective charge distribution (in Fourier space):
\ba
 \frac{1}{e} \hat{K}(\kv, Q) = &-& 
 \frac{\kappa_0^2}{\kappa_{\! R}^2}
\frac{(\kappa_R R_c)^3}{(\kappa_R b)} 
\frac{\left(m_{\! R} n - n_{\! R} m \right)}{mn(m+n)}
\Psi_0(kR_c, \kappa_{\! R}R_c)
\label{K-hat-general}\\
&+& \frac{Q}{e} \, \big[ 1+ \frac{\kappa_{\! R}^2}{k^2}
+ f_2 (kd, \kappa_{\! R}d) 
 \big]
\nonumber\\
&+& \left( \!\frac{Q}{ e}\!\right)^2 (m_{\! R} - n_{\! R})\,
(\kappa_{\! R} b)\,\Psi_2 (kR_c, \kappa_{\! R} R_c),
\nonumber
\ea 
where $\Psi_0, \Psi_2$ are defined in Eqs.~(\ref{Psi_0-def}) and (\ref{Psi_2-def-2}).  

\subsection{Renormalized Debye length of asymmetric electrolytes}
We can now set $Q = m \,e, - n\, e$ in Eq.~(\ref{K-hat-general}) to obtain the effective charge distributions for each specie of constituent ions in the bulk and use Eq.~(\ref{alpha-K-relation}) to find the linear response kernel $\hat{\alpha}(\kv)$.  We shall skip the calculation details and present the results directly.  The kernel $\hat{\alpha}(\kv)$ is 
\ba
\hat{\alpha}(\kv) &=& \kappa_0^2 \Big[ 
1+ \frac{\kappa_{\! R}^2}{k^2}   
+ f_2 (kR_c, \kappa_{\! R}R_c)
\nonumber\\
&+& (m-n) (m_{\! R} - n_{\! R}) (\kappa_{\! R} b) 
\Psi_2(k R_c, \kappa_{\! R} R_c)
\Big]. 
\ea
On the other hand, setting $k = i \kappa_{\! R}$ in Eq.~(\ref{K-hat-general}), and using Eq.~(\ref{G_R-expansion-2-3}), we find the renormalized charge of a hard sphere with bare charge $Q$:
\begin{subequations}
\be
\frac{Q_{\! R}}{e} = a_0 + a_1 \left( \!  \frac{Q}{e} \! \right)
 + a_2 \, \left( \!  \frac{Q}{e} \! \right)^2 + O(Q^3). 
 \label{Q-RG-expansion}
\ee
where the coefficients $a_0, a_1, a_2$ are 
\ba
a_0 &=& 
-  \frac{\kappa_0^2}{\kappa_{\! R}^2}
\frac{1}{(\kappa_{\! R} b)} 
\frac{\left(m_{\! R} n - n_{\! R} m \right)}{mn(m+n)}
\nonumber\\
&\times &\left[ 
- \frac{1}{e} \text{E}_1(- \kappa_{\! R}R_c - 1)
+ \frac{1}{e} \text{E}_1(-1) 
+ e^{\kappa_{\! R}R_c} (\kappa_{\! R}R_c - 2) + 2 
\right],
\nonumber\\
a_1 &=& \frac{e^{\kappa_{\! R}R_c}}{1+\kappa_{\! R}R_c }, 
\nonumber\\
a_2 &=& (m_{\! R} - n_{\! R})\,
(\kappa_{\! R} b) e^{2 \kappa_{\! R}R_c}
   \label{Q_R-general}\\
&\times& 
\left[ \frac{  
{\rm E}_1 (\kappa_{\! R}R_c)}{4 (\kappa_{\! R}R_c+1)^2}
+  \frac{(\kappa_{\! R}R_c -1) e^{2 \kappa_{\! R}R_c}  
{\rm E}_1 (3\kappa_{\! R}R_c)}{4 (\kappa_{\! R}R_c +1)^3}
\right].
 \nonumber
\ea
\end{subequations}
Note that for symmetric electrolytes, the even order coefficients $a_0, a_2$ are contained to vanish by symmetry, since $m_{\! R} = n_{\! R}$, and $m = n$.   The lowest order renormalization is therefore of order $Q^3$, see Eq.~(\ref{kappa_R-sym}).

\begin{figure*}[tbp!]
	\centering
\subfigure[]{	
		\includegraphics[width=5.2cm,height=6.2cm]{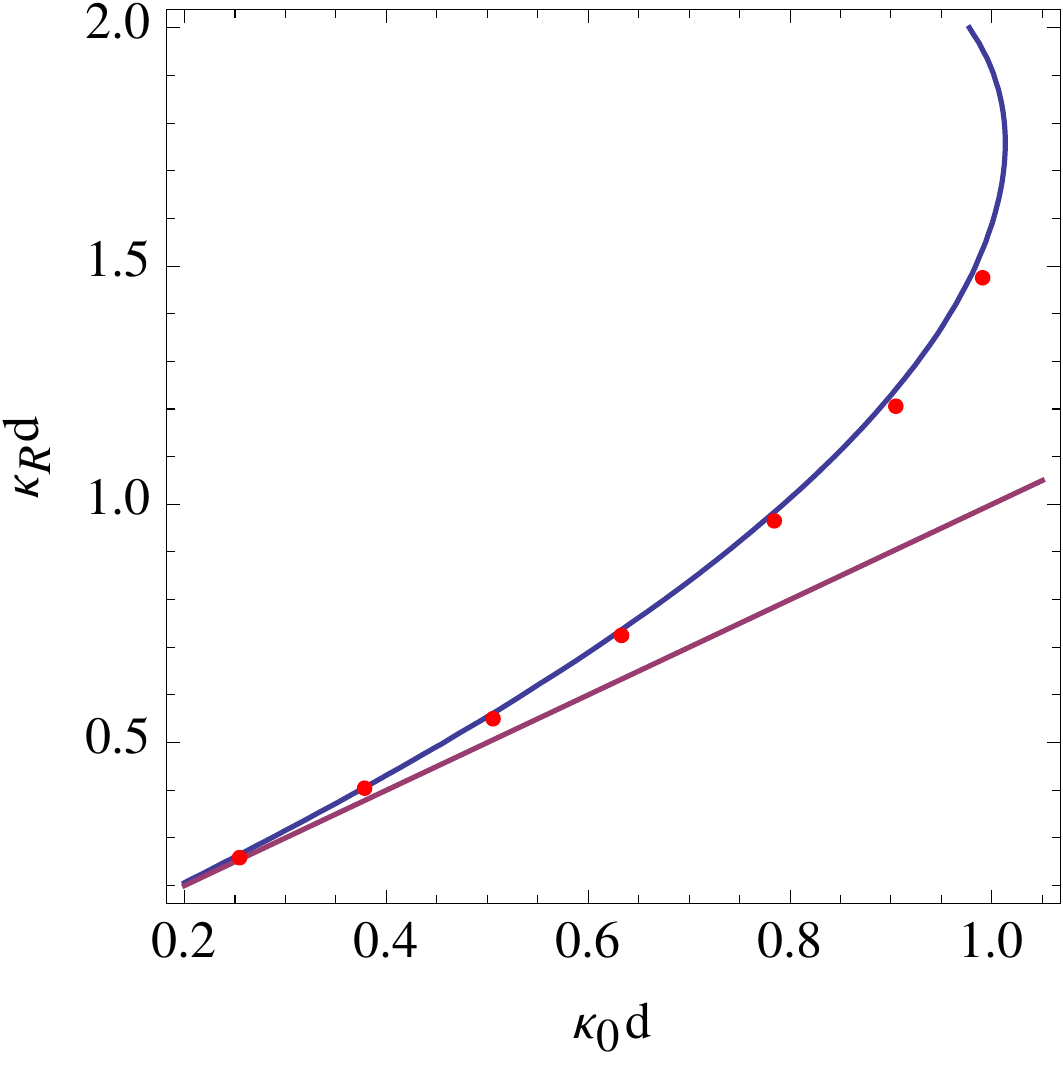}
}
\subfigure[]{	
	\includegraphics[width=5.2cm,height=6.2cm]{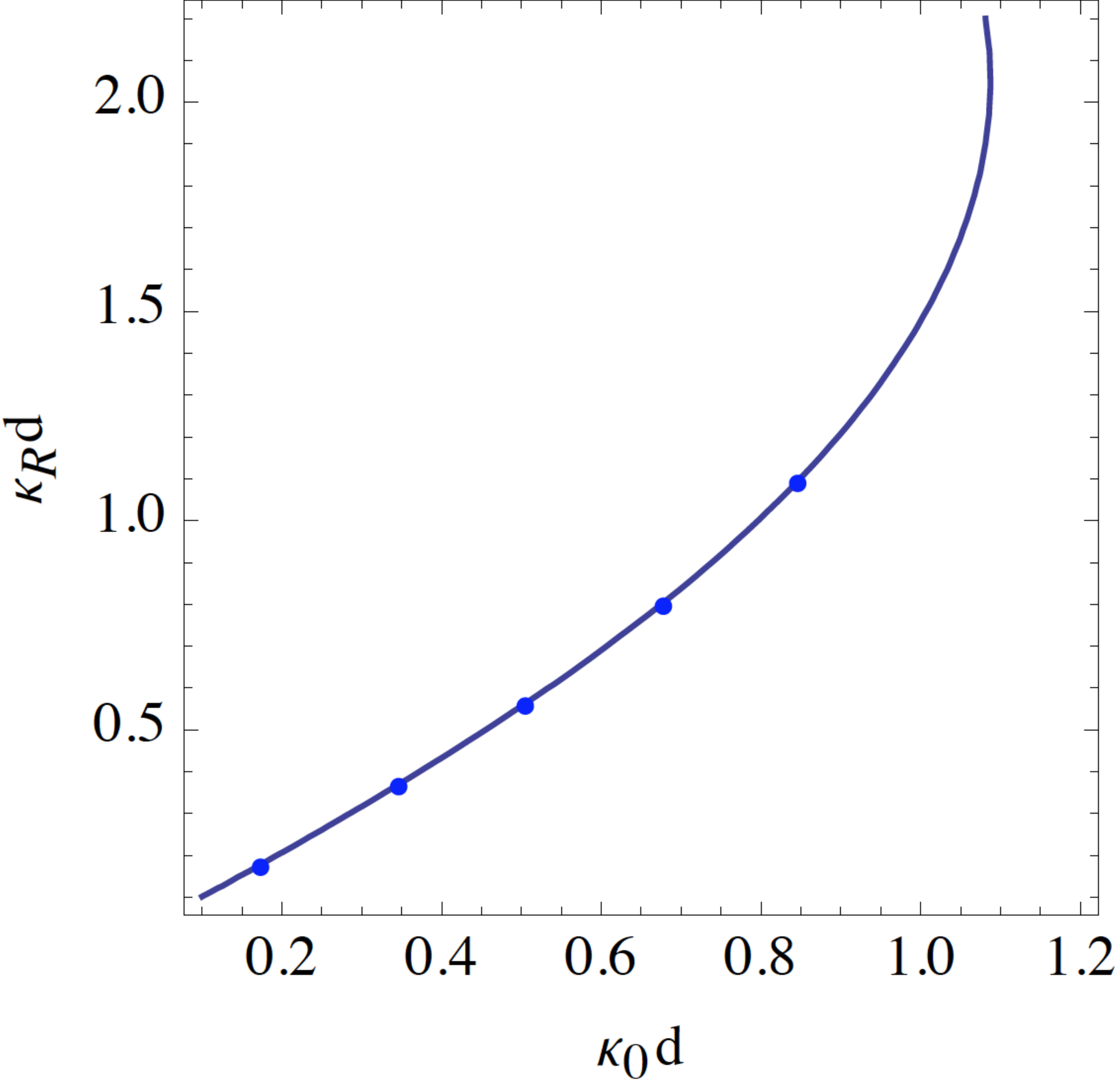}
}
\subfigure[]{	
		\includegraphics[width=5.3cm]{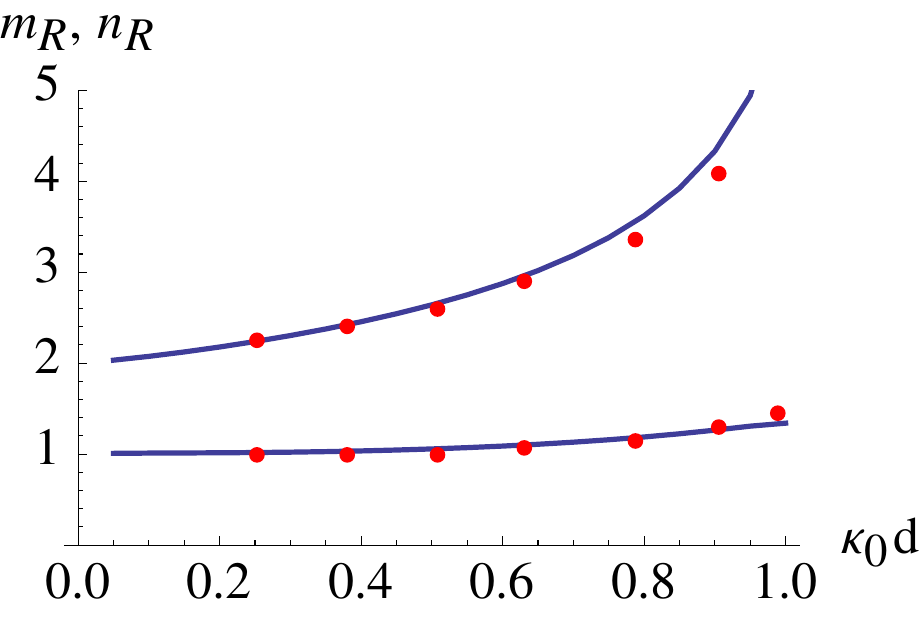} 
}
\subfigure[]{	
	\includegraphics[width=5.3cm]{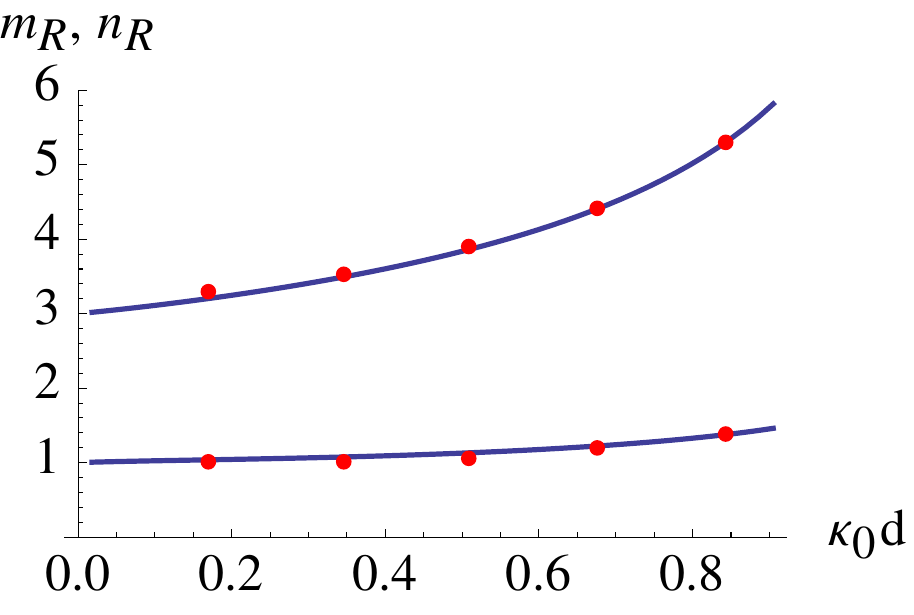} 
}
\subfigure[]{	
		\includegraphics[width=5cm]{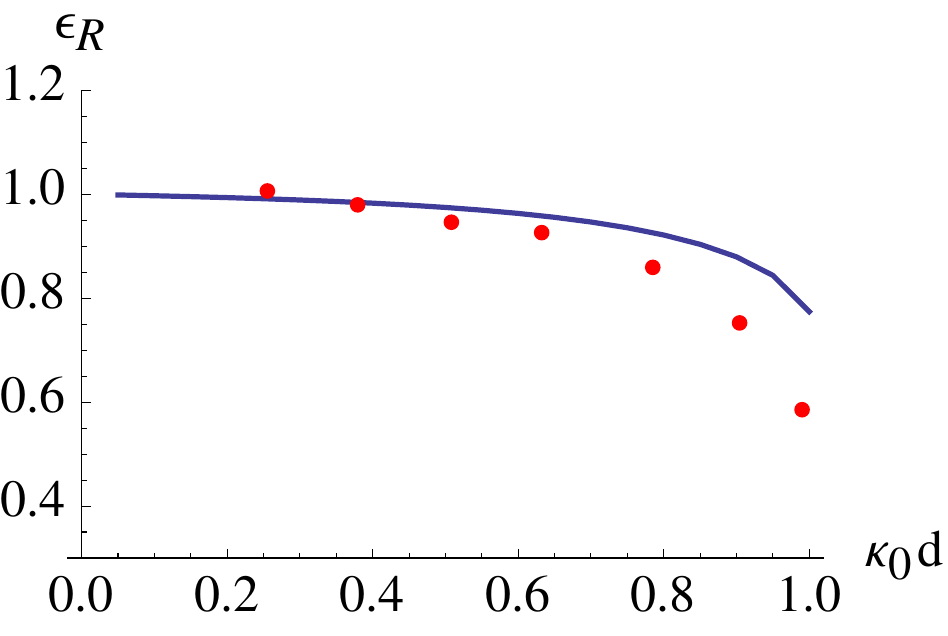} 
}
\subfigure[]{	
	\includegraphics[width=5cm]{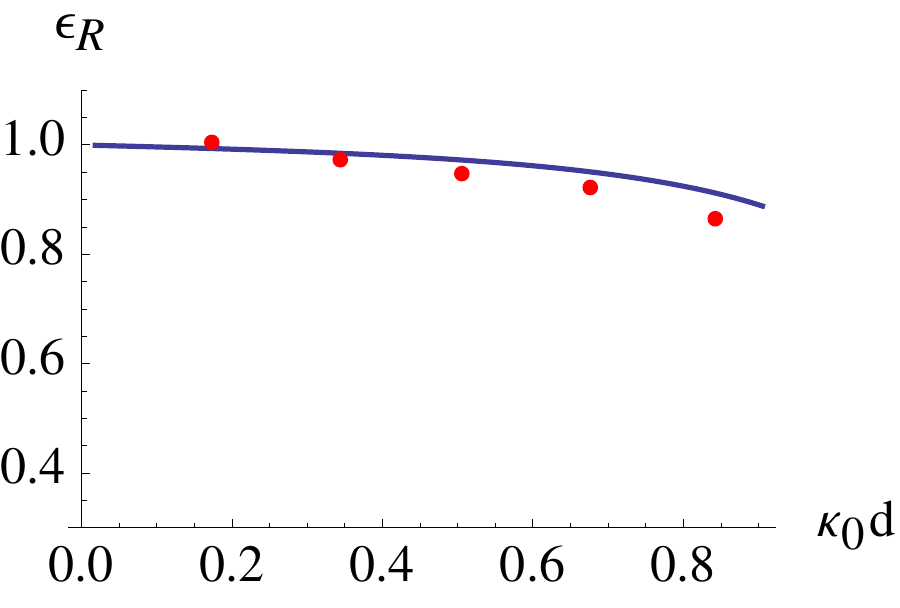}
}
\vspace{-3mm}
\caption{{\bf Renormalized parameters for asymmetric electrolyte.}  Comparison of theoretical predictions Eqs.~(\ref{RG-results-asym}) with large scale MC simulations. Panels (a), (c), (e) :  $2:-1, d = 7.5 \AA$.  The straight-line in panel (a) is the prediction of classical PB.  
Panels (b), (d), (f): $3:-1, d = 24 \AA$.   Note that the threshold densities for charge oscillations for these systems are much lower than that for symmetric electrolytes. 
 Substantial disagreements between theory and simulation for $\epsilon_{\! R}$ arise near the threshold of charge oscillation.  This is probably due to the local approximation Eq.~(\ref{local-approx-alpha}).  Simulations details are discussed in a separate publication \cite{DXL-GPU}. 
\vspace{-3mm} }
  \label{fig:fitting-2-1}
\end{figure*}

Setting $Q = m e, -n e$, the l.h.s. of Eq.~(\ref{Q-RG-expansion}) reduces to the {\em renormalized valences} $m_{\! R} , -n_{\! R}$ of the constituent ions.  Solving for $m_{\! R},n_{\! R}$ we find:
\begin{subequations}
\label{RG-results-asym}
\ba
&& m_R =
 \quad  \label{m_R-m-general}
\\
&& \frac{F_1(\kappa_{\! R}d) m \big[ 
F_0(\kappa_{\! R}d) {\kappa_0^2}/{\kappa_{\! R}^2}  
+ m n s (F_2(\kappa_{\! R}d) s n (m+n)-1)\big]}
{m n s \left[ \, F_2(\kappa_{\! R}d) s
   \left(m^2+n^2\right)-1 \, \right]
   -  \left[\, F_2(\kappa_{\! R}d) s (m-n)^2-1\, \right] 
   F_0(\kappa_{\! R}d){\kappa_0^2}/{\kappa_{\! R}^2} },
\nonumber\\
&& n_R  =
  \label{n_R-n-general}
\\
&& \frac{F_1(\kappa_{\! R}d) n \big[ F_0(\kappa_{\! R}d) 
 {\kappa_0^2}/{\kappa_{\! R}^2}  
+ m n s  (F_2(\kappa_{\! R}d) s m (m+n)-1) \big] }
{m n s \left[\, F_2(\kappa_{\! R}d) s
   \left(m^2+n^2\right)-1 \, \right]
   - \left[\, F_2(\kappa_{\! R}d) s (m-n)^2-1 \, \right] F_0(\kappa_{\! R}d)
    {\kappa_0^2}/{\kappa_{\! R}^2} }.
\quad \nonumber
\ea
where $s = \kappa_{\! R} b$ and functions $F_0(y), F_1(y), F_2(y)$ are defined as
\ba
F_0(y) &=&
 \frac{1}{ y^3} \Big[   
\frac{1}{e} \,  \text{E}_1(-1-y)
 - \frac{1}{e} \, \text{E}_1(1) 
 + e^{y} (2-y) -2
\Big], \nonumber\\
F_1(y) &=&
 \frac{e^{y}}{1+y},
 \label{F-012-def}\\
F_2(y)  &= &
 \frac{e^{2y}}{ 4 (1+ y)^3} \Big[ 
  ( 1 - y ) \, e^{2 y}  {\rm E}_1 (3 y) 
+ (1+ y) {\rm E}_1 (y)
\Big].    \nonumber
\ea
Plugging these back into Eq.~(\ref{kappaR-kappa}), we find a self-consistent equation for $\kappa_{\! R}$:
\ba
&& \left(\! \frac{\kappa_{\! R}}{\kappa_0} \!\right)^2 
=    \label{kappa_R-kappa_0-general}
\\
 && 
 \frac{F_1(\kappa_{\! R}d) 
 \left[
 F_0( \kappa_{\! R} d )  
\left( \! \frac{\kappa_0} { \kappa_{\! R} } \! \right)^2
+ m n s \left[ \, 2 F_2(\kappa_{\! R}d) m n s-1 \, \right]
\right]}
   { m n s \big[  F_2(\kappa_{\! R}d) s  \left(m^2+n^2\right)-1 \big] 
   -   \big[ F_2(\kappa_{\! R}d) s (m-n)^2-1\big] 
   F_0(\kappa_{\! R}d) 
    \left( \! \frac{\kappa_0} {\kappa_{\! R} } \! \right)^2}. 
 \nonumber
 \ea
This gives the renormalized inverse Debye length $\kappa_{\! R}$ as an implicit function of the bare one $\kappa_0$.  Finally the renormalized dielectric constant can also be obtained using Eq.~(\ref{G_R-expansion-2}) (with shorthand $y = \kappa_{\! R} d$ below): 
\ba
\frac{\epsilon_R}{\epsilon} -1 &=&
 -\frac{1}{24 (y+1)^3}  
 \left( \! \frac{\kappa_0} {\kappa_{\! R} } \! \right)^2
 \Bigg\{  \nonumber\\
&&
 12 (y+1)^2 
 \left[\, 2+ 2 y-3 e^y+y \sinh y + (y+1)
   \cosh y \, \right]
 \label{ep_R-general}\\
 &+& s \, e^{-y} (m-n) (m_{\! R} - n_{\! R})
   \Big[ \, (y+1) \left[ \,3 e^{3 y} \text{Ei}(-y) + 3 e^{2 y}+1 \, \right]
 \nonumber\\
&-& 3 e^{3 y} \text{Ei}(-3 y)
 \left[\, 2 e^y  \left( y^2 + 3 \right) \sinh  y
 - 6 y e^y \cosh y  + y + 1 \, \right] \Big]  \Bigg\}.
 \nonumber
 \ea
\end{subequations}
In Fig.~\ref{fig:fitting-2-1}  we show the comparison between the analytic results Eqs.~(\ref{RG-results-asym}) and large scale MC simulations of two systems: 1)  $2:-1, d = 7.5 \AA$, and 2)  $3:-1, d = 24 \AA$.  It can be seen that the agreement between theory and simulations is generally good.  Note that as the density increases, the valence of larger ion is substantially renormalized upwards by ionic correlations, whilst that of smaller ions remains approximately the same.  Furthermore, both systems exhibit charge oscillations in the high density regime. The threshold value $\kappa^*_0 d$ is approximately unity, not too much different from that of symmetric electrolytes.  However, because the bare Debye length is related to ion valences nonlinearly via Eq.~(\ref{alpha-kappa_0}), the threshold ion density for charge oscillation is much lower in asymmetric electrolytes than in asymmetric electrolytes.  
 
Let us now check the special case of symmetric electrolytes, where $m = n, m_{\! R} = n_{\! R}$.  All results reduce to those in Sec.~\ref{sec:sym-electrolyte}.  In particular, Eqs.~(\ref{m_R-m-general})-(\ref{kappa_R-kappa_0-general}) reduce to Eq.~(\ref{kappa_R-sym-1}), and Eq.~(\ref{ep_R-general}) reduces to Eq.~(\ref{ep_R-sym}), as it should be.  

Let us also check the point-like ion limit, where $d \rightarrow 0$.  The renormalized charge Eq.~(\ref{Q-RG-expansion}) reduces to the following limiting form:
\ba
\frac{Q_{\! R}}{e}  =   \frac{Q}{e} 
 + \frac{\log 3}{4} (\kappa_{\! R} b) (m_{\! R} - n_{\! R})
 \left( \!  \frac{Q}{e} \! \right)^2 + O(Q^3). 
 \label{Q-RG-expansion-1}
\ea
Hence charge renormalization becomes significant if the renormalized Debye length becomes comparable with the Bjerrum length, i.e., when $(\kappa_{\! R} b) \sim 1$.
In the same limit, Eqs.~(\ref{F-012-def}) reduce to $F_0 \rightarrow 0, F_1 \rightarrow 1, F_2 \rightarrow \frac{1}{4} \log 3$, and hence Eqs.~(\ref{m_R-m-general}) and (\ref{n_R-n-general})  reduce to 
\ba
\frac{m_R}{m} &=& \frac{4 - n (\kappa_{\! R}b)   (m+n) \log (3)}
{4- (\kappa_{\! R}b)  \left(m^2+n^2\right) \log (3) }, \\
\frac{n_R}{n} &=& \frac{4 - m (\kappa_{\! R}b)  (m+n)\log (3) }
{4- (\kappa_{\! R}b)  \left(m^2+n^2\right)\log (3) }.
\ea
Eq.~(\ref{kappa_R-kappa_0-general}) reduces to
\be
\frac{\kappa_{\! R}^2}{\kappa_0^2} = 
\frac{4 - 2 (\kappa_{\! R}b) m n \log (3)}
{4- (\kappa_{\! R}b) \left(m^2+n^2\right) \log (3) },
\ee
and finally Eq.~(\ref{ep_R-general}) reduces to 
\be
\frac{\epsilon_R}{\epsilon}  = 
1 - \frac{ (\kappa_{\! R}b)  ( \, 4-3 \log 3 \, ) 
(m-n)^2}{12 (2-m n  (\kappa_{\! R}b)  \log 3 \, )}. 
\ee
These results also predict charge oscillation if $(\kappa_{\! R}b) $ is comparable with unity, which clearly indicate that charge oscillation in asymmetric electrolytes can be solely driven by electrostatic correlations (mainly between ions of higher valences). 

Strictly speaking, a two-component plasma model with point-like ions is not well-defined 
because of the instability towards annihilation of opposite charges.  For small but finite ion sizes, this instability is manifested as formation of bound pairs of ions, or even larger clusters.  This instability does not show up at the level of our approximation, just as in the classical Debye-Huckel theory.  Linearization, which is adopted in both theories, is responsible for the suppression of this instability at short scales.  As a logical consequence, whenever bound ion clusters can not be ignored, linearization breaks down, and the short scale properties derived in our theory (similar to those of Debye-Huckel theory) can not be trusted.  This happens, for example, for dense electrolytes with small ions where the electrostatic energy between neighboring ions becomes much larger than the thermal energy $k_B T$.  Indeed, we have also simulated $3:-1$ electrolyte with $d = 7.5 \AA$.  The largest interaction energy between two ions (in close contact) is then approximately $3 k_B T$, which makes linearization a bad approximation in the dense regime.  As expected, we found that substantial disagreement between theory and simulation.  

\section{Conclusion and Acknowledgement}
Our results demonstrate that renormalization of ion valences and Debye length, dielectric constant is more apparent in asymmetric electrolytes than in symmetric.  We also find that, generically, the valence of larger ions is renormalized substantially upwards, whereas that of smaller ions remains stays approximately constant.  Finally, the threshold density for charge oscillation in asymmetric electrolytes is much lower than that for symmetric electrolytes. 

We thank NSFC (Grants No. 11174196 and 91130012) for financial support.  We also thank  Wei Cai for interesting discussions.






\begin{thebibliography}{10}
\bibitem{Kirkwood:1936ys}
John~G Kirkwood.
\newblock Statistical mechanics of liquid solutions.
\newblock {\em Chemical Reviews}, 19(3):275--307, 1936.

\bibitem{Reviews:OCP}
Baus, Marc, and Jean-Pierre Hansen. "Statistical mechanics of simple Coulomb systems." 
Physics Reports 59.1 (1980): 1-94. \\
http://www.sciencedirect.com/science/article/pii/0370157380900228

\bibitem{Reviews:charge oscillation}
More references on charge oscillations

\bibitem{Hansen:2013qd}
Jean-Pierre Hansen and Ian R. McDonald.
\newblock {\em Theory of simple liquids: with applications to soft matter}.
\newblock Academic Press, 2013.


\bibitem{lee1997charge}
Benjamin~P Lee and Michael~E Fisher.
\newblock Charge oscillations in debye-h{\"u}ckel theory.
\newblock {\em EPL (Europhysics Letters)}, 39(6):611, 1997.


\bibitem{DLX-ion-specific}
Mingnan Ding, Yihao Liang, and Xiangjun Xing, Surfaces with Ion-specific Interactions, Their Effective Charge Distributions and Effective Interactions, to be submitted. 










\bibitem{Onsager:1933lq}
Lars Onsager.
\newblock Theories of concentrated electrolytes.
\newblock {\em Chemical Reviews}, 13(1):73--89, 1933.

\bibitem{Kjellander:1992nr}
Roland Kjellander and D~John Mitchell.
\newblock An exact but linear and poisson---boltzmann-like theory for
  electrolytes and colloid dispersions in the primitive model.
\newblock {\em Chemical physics letters}, 200(1):76--82, 1992.

\bibitem{Kjellander:1994xy}
Roland Kjellander and D.~John Mitchell.
\newblock Dressed ion theory for electrolyte solutions: A
  Debye-H{\"u}ckel-like reformulation of the exact theory for the primitive
  model.
\newblock {\em The Journal of Chemical Physics}, 101(1):603--626, 1994.

\bibitem{Ennis:1995qf}
Jonathan Ennis, Roland Kjellander, and D.~John Mitchell.
\newblock Dressed ion theory for bulk symmetric electrolytes in the restricted
  primitive model.
\newblock {\em Journal of Chemical Physics}, 102(2):975, 01 1995.

\bibitem{DIT-review-Kjellander} 
Roland Kjellander.
\newblock ``Distribution Function Theory of Electrolytes and Electrical Double Layers: Charge Renormalisation and Dressed Ion Theory'' 
\newblock in Electrostatic Effects in Soft Matter and Biophysics (C. Holm, P. Kékicheff and R Podgornik, Eds; 
\newblock NATO Science Series, Kluwer Academic Publishers, Dordrecht 2001) pp. 317 – 364.

\bibitem{DIT-review-2003}
Varela, Luis M., Manuel Garcı́a, and Vı́ctor Mosquera. 
\newblock "Exact mean-field theory of ionic solutions: non-Debye screening." 
\newblock {\em Physics reports} 382.1 (2003): 1-111.





  



\bibitem{Stell:1968fk}
G.~Stell and J.~L. Lebowitz.
\newblock Equilibrium properties of a system of charged particles.
\newblock {\em The Journal of Chemical Physics}, 49(8):3706--3717, 1968.

\bibitem{Mitchell:1968fk}
D.~J. Mitchell and B.~W. Ninham.
\newblock Asymptotic behavior of the pair distribution function of a classical electron gas.
\newblock {\em Physical Review}, 174(1):280--289, 10 1968.

\bibitem{0295-5075-12-5-016}
P.~K{\'e}kicheff and B.~W. Ninham.
\newblock The double-layer interaction in asymmetric electrolytes.
\newblock {\em EPL (Europhysics Letters)}, 12(5):471, 1990.

\bibitem{DXL-GPU}
Yihao Liang, Xiangjun Xing, and Yaohang Li. 
\newblock ``A GPU-based Large-scale  Monte Carlo Simulation Method for Systems with Long-range Interactions''.
\newblock submitted to Journal of Computational Physics. 

 



\bibitem{Henderson:1979kq}
Douglas Henderson, Lesser Blum, and Joel~L Lebowitz.
\newblock An exact formula for the contact value of the density profile of a system of charged hard spheres near a charged wall.
\newblock {\em Journal of Electroanalytical Chemistry and Interfacial  Electrochemistry}, 102(3):315--319, 1979.


\bibitem{Debye-charging}
P. W. Debye and E. Huckel. 
\newblock Phys. Z. 24, 185 (1923).

\bibitem{Stone-Math}
M.~Stone and Paul~M. Goldbart.
\newblock {\em Mathematics for Physics}.
\newblock Cambridge University Press, 2009.

\bibitem{Kjellander:1995fk}
Kjellander, Roland. 
\newblock ``Modified Debye-Hückel approximation with effective charges: an application of dressed ion theory for electrolyte solutions.''
\newblock {\em The Journal of Physical Chemistry} 99.25 (1995): 10392-10407.




\bibitem{HMF-Milton-Stegun} 
 {Abramowitz, Milton; Irene Stegun (1964). Handbook of Mathematical Functions with Formulas, Graphs, and Mathematical Tables. Abramowitz and Stegun. New York: Dover. ISBN 0-486-61272-4., Chapter 5}
 
\end{thebibliography}

\end{document}